\documentclass[12pt]{article}
\usepackage{MyArticle}
\usepackage{amsmath,amssymb,amsthm,mathabx,mathrsfs,mathtools,bm,algorithm,algorithmic,graphicx, subfigure} 
\usepackage{enumerate,array,fontenc,multirow,natbib,makecell,booktabs,hhline,chngcntr,dsfont,setspace,rotating,xr,xcolor}
\usepackage[colorlinks,linkcolor=red,anchorcolor=blue,citecolor=blue]{hyperref}

\usepackage[total={6.35in,8.55in}]{geometry}
\allowdisplaybreaks 

\newcommand{\sEM}{\rm{sEM}}

\begin{document}

\title{\Large{\textbf{Statistical Inference for High-Dimensional \\
Vector Autoregression with Measurement Error \\}}}
\author{
\bigskip
Xiang Lyu$^\dag$, Jian Kang$^\ddag$, and Lexin Li$^\dag$ \\ 
\normalsize{\textit{$^\dag$University of California at Berkeley, and $^\ddag$University of Michigan}} \\
}
\date{}

\maketitle

\begin{abstract} 
High-dimensional vector autoregression with measurement error is frequently encountered in a large variety of scientific and business applications. In this article, we study statistical inference of the transition matrix under this model. While there has been a large body of literature studying sparse estimation of the transition matrix, there is a paucity of inference solutions, especially in the high-dimensional scenario. We develop inferential procedures for both the global and simultaneous testing of the transition matrix. We first develop a new sparse expectation-maximization algorithm to estimate the model parameters, and carefully characterize their estimation precisions. We then construct a Gaussian matrix, after proper bias and variance corrections, from which we derive the test statistics. Finally, we develop the testing procedures and establish their asymptotic guarantees. We study the finite-sample performance of our tests through intensive simulations, and illustrate with a brain connectivity analysis example. 
\end{abstract}

\vspace{0.2in}

\newpage
\baselineskip=21pt

\section{Introduction}	
\label{sec: introduction}

In this article, we study statistical inference for high-dimensional vector autoregression (VAR) with measurement error. More specifically, we consider the model, 
\begin{align} \label{eq: model_measure}
\begin{split}
\yb_{t} &=  \xb_{t} + \bepsilon_{t},   \\ 
\xb_{t+1} & =  \Ab_* \xb_{t}  + \bm{\eta}_{t},
\end{split}
\end{align}
where $\yb_{t} = (y_{t,1}, \ldots, y_{t,p})^\top \in \RR^{p}$ is the observed multivariate time series, $\xb_{t} = (x_{t,1}, \ldots,$ $x_{t,p})^\top \in  \RR^{p}$ is the multivariate latent signal that admits an autoregressive structure, $\bepsilon_{t} = (\epsilon_{t,1}, \ldots, \epsilon_{t,p})^\top \in \RR^{p}$ is the measurement error for the observed time series, $\bm{\eta}_{t} = (\eta_{t,1}, \ldots, \eta_{t,p})^\top$ $\in \RR^{p}$ is the white noise of the latent signal, and $\Ab_*= (A_{*,ij})  \in \RR^{p\times p}$ is the sparse transition matrix that encodes the directional relations among the latent signal variables of $\xb_{t}$. Furthermore, we focus on the scenario $\|\Ab_*\|_2 <1$ such that the VAR model of $\xb_{t}$ is stationary. The error terms $\bepsilon_{t}$ and $\bm{\eta}_{t}$ are i.i.d.\ multivariate normal with mean zero and covariance $\sigma_{\epsilon,*}^2 \Ib_p$ and $\sigma_{\eta,*}^2 \Ib_p$, respectively, and are independent of $\xb_{t}$. Here we focus on the lag-1 autoregressive structure and homoscedastic errors. We later discuss potential extensions in Section \ref{sec: discussion}. 

Model like \eqref{eq: model_measure} is frequently employed in a variety of scientific and business applications, e.g., finance, engineering and neuroscience. Our motivation is brain effective connectivity analysis based on functional magnetic resonance imaging (fMRI). The brain is a highly interconnected dynamic system, in which the activity and temporal evolution of neural elements are triggered and influenced by the activities of other elements \citep{Garg2011}. Of great interest in neuroscience is to understand the directional relations among the neural elements through fMRI, which measures synchronized blood oxygen level dependent signals at different brain locations. VAR model is an important tool to model such directional relations, which are encoded by the transition matrix $\Ab_*$, while the stationarity is often assumed \citep{Bullmore2009, Chen2011}. However, unlike a typical VAR, the observed time series $\yb_{t}$ is the contaminated version of the true signal $\xb_{t}$, added with a measurement error $\bepsilon_{t}$ \citep{Zhang2015, Luo2019}. 

We address the statistical inference problem of the transition matrix $\Ab_*$ under model \eqref{eq: model_measure}, and we aim at a high-dimensional setting where  $p^2$ exceeds  the length of series $T$. We first test the global hypotheses,
\begin{equation} \label{eq: global_hypo}
H_{0}: A_{*,ij} = A_{0,ij}, \  \textrm{ for all } (i,j) \in \cS \quad \textrm{versus} \quad  H_{1}: A_{*,ij} \neq A_{0,ij},  \ \textrm{ for some } (i,j) \in \cS,
\end{equation}
for a given $\Ab_{0} = (A_{0,ij}) \in \RR^{p \times p}$ and $\cS \subseteq [p] \times [p]$, where $[p] = \{1, \ldots, p\}$. The most common choice is $\Ab_0=\mathbf{0}_{p\times p}$ and $\cS =[p] \times [p]$. We next test the simultaneous hypotheses, 
\begin{equation}\label{eq: multi_hypo}
H_{0; ij}: A_{*,ij} = A_{0,ij},  \quad \textrm{versus} \quad  H_{1; ij}: A_{*,ij} \ne A_{0,ij},  \ \textrm{ for all } (i, j) \in \cS. 
\end{equation}

There has been a large body of literature studying sparse estimation of $\Ab_*$ in VAR models \citep[among others]{HHC08, SB11, NW11, BM15, HLL15}. However, they all assumed that there is no measurement error $\bepsilon_{t}$, or equivalently, that $\xb_{t}$ is fully observed. Moreover, while both estimation and inference can produce a sparse representation of $\Ab_*$, they are utterly different problems. Sparse estimation  usually does not explicitly control the false discovery rate (type I error), and does not produce an explicit significance quantification ($p$-value). There has been a relative paucity of inference methods for $\Ab_*$ in VAR models. Existing inference solutions mostly focused on the low-dimensional VAR setting; see \cite{Reinsel_book, tsay_book, S15} for a review. More recently, for the high-dimensional VAR setting, \cite{KKP18} proposed to bootstrap the de-biased Lasso estimator, while \cite{ZC19} extended the de-correlated score test of \cite{NL17}. However, they only addressed the global testing problem \eqref{eq: global_hypo}, but not the simultaneous testing problem \eqref{eq: multi_hypo}. Besides, it is unclear how to adapt their tests to accommodate additional measurement error. To the best of our knowledge, there is no existing solution to directly address both global and simultaneous testing problems under the high-dimensional VAR setting with error. 

Our proposal is built upon two key ingredients: a sparse expectation-maximization (EM) algorithm, and the high-dimensional covariance inference. The first ingredient, the EM algorithm, offers a way to estimate model parameters in the existence of measurement error. 
Early EM methods, however, only justified the convergence to a local optimum and did not consider sparsity. Recently, a seminal work of \citet{BWY17} provided sufficient conditions to guarantee the convergence of standard EM to the global optimum but only in a low-dimensional setting, while \cite{CMZ19} extended the guarantee to a high-dimensional sparse Gaussian mixture model. See also \citet{WGNL15, YC15}. On the other hand, they all worked with i.i.d.\ observations, whereas our problem involves temporally highly dependent data. Extension from independent to dependent observations is far from trivial.  The second ingredient, the high-dimensional covariance inference, has been intensively studied in recent years, including both global testing \citep{CJ11, XW13, CZZ10} and simultaneous testing \citep{L13, Cai2013}. See also \citet{C17, CaiSun2017} for reviews. However, they all assumed the data which the covariance is constructed from are fully observed. By contrast, our inference is about the transition matrix $\Ab_*$ of the latent unobserved $\xb_t$, and the covariance of the observed $\yb_t$ is a nonlinear transformation of $\Ab_*$, making it difficult to trace back to $\Ab_*$. Consequently, there is a considerable gap before we can apply the existing covariance inference tools to our setting.

In this article, we develop inferential procedures for both the global and simultaneous testing problems \eqref{eq: global_hypo} and \eqref{eq: multi_hypo} under the high-dimensional VAR model with error. Our proposal includes three main steps. First, we develop a new sparse EM algorithm to estimate relevant model parameters. Next, we construct a Gaussian matrix on the domain of transition matrix, from which we derive the test statistics. Finally, we develop the global and simultaneous testing procedures with proper theoretical guarantees.

In the first step, we develop a new sparse EM algorithm to estimate both the transition matrix $\Ab_*$ and the error variances $\sigma_{\epsilon,*}^2$ and $\sigma_{\eta,*}^2$. In particular, the maximization step is done via a generalized Dantzig selector for Yule-Walker equation, which can be efficiently solved by parallel linear programming \citep{CT07, HLL15}. We then establish the convergence of our sparse EM estimators to the true parameters, within the statistical precision required for the test statistics and the transition matrix inferences in later steps. We note that, the existing EM theory adopts the log-likelihood in an infinite-sample scheme as the key analytical tool, which becomes an expectation at a single observation given i.i.d.\ observations \citep{BWY17, CMZ19}. However, the temporal dependence in our model makes the expectation of the log-likelihood change with the sample size. To tackle the issue, in our theoretical analysis, we consider the expectation in a finite-sample scheme instead, which introduces additional technical difficulty. We then derive several new concentration inequalities to establish the statistical error under some weak sparsity assumptions.  

In the second step, we construct a Gaussian matrix as the test statistic for our transition matrix inference. This is built on a key observation that the inference on $\Ab_*$ is equivalent to the inference on the lagged auto-covariance of some noise term. Since this noise is not directly observed, we employ the sparse EM algorithm in the first step to reconstruct the noise. We then study the non-asymptotic behavior of the sample lagged auto-covariance of the reconstructed noise, and explicitly characterize its bias and variance. This in turn leads to the construction of the test statistic matrix whose entries marginally follow a standard Gaussian distribution under the null hypothesis.

In the third step, we develop a global testing procedure based on the extreme distribution of the maximal entry of the test statistic matrix from the second step, and develop a simultaneous testing procedure by thresholding at a level that controls false discovery rate (FDR). Theoretically, we obtain the asymptotic size and power of the global test, which together establish the consistency of our test. We also show that our simultaneous test achieves a consistent FDR control. Our testing procedures are extensions of the covariance inference methods such as \citet{CJ11, L13, Cai2013}. But unlike the existing methods that are built on the sample covariance of fully observed data, our tests are obtained from the sample lagged auto-covariance of the reconstructed noise. This difference requires us to derive new concentration inequalities and Gaussian approximations to disentangle the reconstruction error, lag effect, and temporal dependence. These new theoretical results themselves may be of independent interest. 

We employ the following notation throughout this article. Let $|\cS|$ denote the cardinality of a set $\cS$. For a scalar $a \in \RR$, let $\lceil a \rceil$ and $\lfloor a \rfloor$ denote the smallest and largest integer greater than or smaller than $a$. For two scalars $a, b \in \RR$, let $a \vee b $ and $a \wedge b$ denote the maxima and minima. For a vector $\mathbf{a}= (a_1 ,\ldots , a_p)^\top \in \mathbb{R}^{p}$, define $\|\mathbf{a}\|_1 = \sum_{i=1}^{p} |a_i|$, $\|\mathbf{a}\|_2 = (\sum_{i=1}^{p} |a_i|^2)^{1/2}$,  and $\|\mathbf{a}\|_{\infty}=\max_{1 \le i \le p} |a_i|$. For an index set $\cS \subseteq [p]$, let $\ab_\cS$ denote the sub-vector of $\ab$ containing only the coordinates indexed by $\cS$. For a matrix $\Mb=(M_{ij})  \in \RR^{p_1 \times p_2}$, define $\| \Mb \|_1 = \sum_{ij} |M_{ij}|$, $\|\Mb\|_2 = \lambda^{1/2}_{\max} (\Mb^\top \Mb) $, $\| \Mb \|_{F} = (\sum_{ij} M_{ij}^2)^{1/2}$, $\|\Mb\|_{\max}= \max_{ij} |M_{ij}|$, $\|\Mb\|_{l_1}=\max_{j \in [p_2]} \sum_{i=1}^{p_1}|M_{ij} |$, $\|\Mb\|_{l_\infty}=\max_{i\in [p_1]} \sum_{j=1}^{p_2}|M_{ij} |$, $\|\Mb\|_{l_\infty}=\max_{i\in [p_1]}$ $\sum_{j=1}^{p_2}|M_{ij}|$, and $\|\Mb\|_{r,2}=\max_{i\in [p_1]} \sqrt{\sum_{j=1}^{p_2}|M_{ij} |^2}$ to be its element-wise $\ell_1$ norm, spectral norm, Frobenius norm, max norm, maximum absolute column sum, maximum absolute row sum, and maximal row-wise Euclidean norm, respectively. Let $\Mb_{i:}$ and $\Mb_{:j}$ denote the $i$th row and $j$th column. Let $\lambda_{\min} (\Mb)$ and $\lambda_{\max}(\Mb)$ denote its smallest and the largest eigenvalue, $\tr(\Mb)$ the trace, and $|\Mb|$ the determinant. Define $D(\Mb)$ as a diagonal matrix whose diagonal elements are the same as those of $\Mb$. 

The rest of the article is organized as follows. Section \ref{sec: EM} presents the sparse EM algorithm, Section \ref{sec: test-statistics} constructs the test statistic matrix, and Section \ref{sec: inference} develops the global and simultaneous testing procedures. Section \ref{sec: simulations} presents the simulations, and Section \ref{sec: HCP} illustrates with a brain connectivity analysis example. Section \ref{sec: discussion} concludes the paper with a discussion. All proofs are relegated to the Supplementary Appendix.

\section{Sparse EM Estimation} 
\label{sec: EM}

\subsection{Sparse EM algorithm}

Let $\{ \yb_{t},\xb_{t} \}_{t=1}^{T}$ denote the complete data, where $T$ is the total number of observations, $\yb_{t}$ is observed but $\xb_{t}$ is latent. Let $\Theta = \left\{ \Ab, \sigma_{\eta}^2, \sigma_{\epsilon}^2 \right\}$ collect all the parameters of interest in model \eqref{eq: model_measure}, and $\Theta_* = \left\{ \Ab_*, \sigma_{\eta,*}^2, \sigma_{\epsilon,*}^2 \right\}$ denote the true parameters. 
The goal is to estimate $\Theta_*$ by maximizing the log-likelihood function of the observed data, $\ell (\Theta | \{\yb_{t}\}_{t=1}^T)$, with respect to $\Theta$.  The computation of $\ell (\Theta | \{\yb_{t}\}_{t=1}^T)$, however, is highly nontrivial. The standard EM algorithm then turns to an auxiliary function, named the finite-sample $Q$-function, 
\begin{eqnarray*} \label{eqn:Q-function}
Q_y (\Theta | \Theta') = \EE \left[ \ell\left( \Theta | \{ \yb_{t},\xb_{t} \}_{t=1}^{T} \right) | \{ \yb_{t}\}_{t=1}^T, \Theta' \right],
\end{eqnarray*} 
which is defined as the expectation of the log-likelihood function for the complete data $\ell(\Theta | \{ \yb_{t},\xb_{t} \}_{t=1}^{T})$, conditioning on a parameter set $\Theta'$ and the observed data $\yb_t$, and the expectation is taken with respect to the latent data $\xb_t$.  The $Q$-function can be computed efficiently, and provides a lower bound of the target log-likelihood function $\ell (\Theta|\{\yb_{t}\}_{t=1}^T)$ for any $\Theta$. The equality $\ell (\Theta'|\{\yb_{t}\}_{t=1}^T) = Q_y(\Theta' | \Theta')$ holds if $\Theta = \Theta'$. 
Maximizing Q-function provides an uphill step of the likelihood. Starting from an initial set of parameters $\hat\Theta_0$, the EM algorithm then alternates between the expectation step (E-step), where the $Q$-function $Q_y (\Theta | \hat\Theta_{k})$  conditioning on the parameters $\hat\Theta_{k}$ of the $k$th iteration is computed, and the maximization step (M-step), where the parameters are updated by maximizing the $Q$-function $\hat\Theta_{k+1} = \argmax_{\Theta} Q_y (\Theta | \hat\Theta_{k})$. 
 
For our problem, we carry out the E-step via the standard Kalman filter and smoother \citep{GH96}. For the M-step, the maximizer of $Q_y(\Theta | \hat{\Theta}_{k})$ satisfies that $(T-1)^{-1} \sum_{t=1}^{T-1} \Eb_{t,t+1;k} = \{(T-1)^{-1} \sum_{t=1}^{T-1} \Eb_{t,t;k} \}\Ab^\top$, where $\Eb_{t,s;k} = \EE \left\{ \xb_{t}\xb_{s}^\top | \{\yb_{t'}\}_{t'=1}^T, \hat{\Theta}_{k-1} \right\}$ for $s, t\in [T]$ is obtained from the E-step. The standard EM algorithm directly inverts the matrix involving $\Eb_{t,t;k}$'s, which is computationally challenging when the dimension $p$ is high. In addition, it yields a dense estimator of $\Ab_*$, leading to a divergent statistical error. To overcome these issues, we propose a sparse EM algorithm to deal with the high dimensionality and to produce a sparse estimate of the transition matrix. Specifically, we consider a generalized Dantzig selector for Yule-Walker equation \citep{CT07, HLL15}, 
\begin{equation} \label{eq: sparse_A}
\hat{\Ab}_{k} = \argmin_{\Ab \in \RR^{p\times p}}  \|\Ab\|_1, \;\; \textrm{such that} \;  \left\| \frac{1}{T-1} \sum_{t=1}^{T-1}   \Eb_{t,t+1;k} -\frac{1}{T-1} \sum_{t=1}^{T-1} \Eb_{t,t;k} \Ab^\top \right\|_{\max} \le \tau_k,
\end{equation}
where $\tau_k$ is the tolerance parameter that is tuned via cross-validation.
 The optimization problem \eqref{eq: sparse_A} is solved using linear programming in a row-by-row parallel fashion. We next update the variance estimates as, 
\begin{align} \label{eqn: epsilon}
\begin{split}
\hat\sigma_{\eta,k}^2 & =  \frac{1}{p(T-1)} \sum_{t=1}^{T-1} \left\{ \tr( \Eb_{t+1,t+1;k})  -   \tr\left ( \hat{\Ab}_{k}  \Eb_{t,t+1;k} \right) \right\} , \\
\hat\sigma^2_{\epsilon,k} & =  \frac{1}{pT } \sum_{t=1}^{T} \left\{ \yb_{t}^\top \yb_{t} - 2 \yb_{t}^\top  \Eb_{t;k} + \tr (\Eb_{t,t;k}) \right\}, 
\end{split}  
\end{align} 
where $\Eb_{t;k} = \EE \{ \xb_{t} | \{\yb_{t'}\}_{t'=1}^T, \hat{\Theta}_{k-1} \}$ for $t \in [T]$, and \eqref{eqn: epsilon} comes from taking derivative on $Q_y(\Theta | \hat{\Theta}_{k})$. We terminate our  sparse EM algorithm when the estimates are close enough in two consecutive iterations, e.g., $\min \left\{ \|\hat{\Ab}_{k} -\hat{\Ab}_{k-1} \|_F ,  | \hat{\sigma}_{\eta,k }-\hat{\sigma}_{\eta,k-1}| ,| \hat{\sigma}_{\epsilon, k}-\hat{\sigma}_{\epsilon, k-1}| \right\} \le 10^{-3}$.

We summarize our sparse EM procedure in Algorithm \ref{alg: sparseEM}. 

\begin{algorithm}[t!]
\caption{Sparse EM algorithm for parameter estimation in model \eqref{eq: model_measure}.} 
\label{alg: sparseEM}
\begin{algorithmic}
\STATE Initialization: $\hat\Theta_0 = \left\{ \hat{\Ab}_{0}, \hat{\sigma}_{\eta,0}^2, \hat{\sigma}_{\epsilon,0}^2 \right\}$, and set $k=1$. 
\REPEAT 
\STATE 1. E-step: Obtain $\Eb_{t;k}$, $\Eb_{t,t;k}$, and $\Eb_{t,t+1;k}$ via Kalman filter and smoothing, conditional on $\{\yb_{t} \}_{t\in[T]}$ and $\hat\Theta_{k-1}$. 
\STATE 2. M-step: 
\STATE \ \ \ \ 2.1. Compute $\hat{\Ab}_{k}$ by \eqref{eq: sparse_A}. 
\STATE \ \ \ \ 2.2. Compute $\hat{\sigma}_{\eta,k}^2$ and $\hat{\sigma}_{\epsilon, k}^2$ by \eqref{eqn: epsilon}. 
\STATE 3. Collect $\hat\Theta_{k} = \left\{ \hat{\Ab}_{k}, \hat{\sigma}_{\eta,k}^2, \hat{\sigma}_{\epsilon,k}^2 \right\}$, and set $k = k + 1$. 
\UNTIL{the stopping criterion is met.}
\end{algorithmic}
\end{algorithm}

\subsection{Estimation consistency}

We next establish the estimation consistency of our sparse EM estimators, and show that they achieve estimation errors within statistical precision required for the construction of test statistic and inference methods in subsequent steps. 

Similar to \citet{BWY17, CMZ19}, we first introduce a concept, the population $Q$-function, as
\begin{eqnarray*} \label{eqn:Q-function-population}
Q (\Theta | \Theta') = \EE \left\{ Q_y(\Theta | \Theta' ) | \Theta_* \right\},
\end{eqnarray*} 
where the expectation is with respect to the observed data $\yb_t$, and this population $Q$-function depends on the true parameters $\Theta_*$ and the sample size $T$. On the other hand, our definition is not exactly the same as that of \citet{BWY17, CMZ19}. They considered the limit of infinite i.i.d.\ observations, which naturally leads to the expectation of $Q_y$ at a single observation by the law of large numbers. However, the temporal dependence in our problem makes the expectation of $Q_y$ change with the sample size $T$.  So we define the population $Q$-function as the expectation at a finite $T$. This change brings additional technical difficulty for  subsequent theoretical analysis. 

We next introduce a sequence of intermediate estimators, $\Theta_{k+1} = \argmax_{\Theta} Q (\Theta | \hat\Theta_{k})$. Note that $\Theta_{k+1}$ is obtained by maximizing the population $Q$-function $Q (\cdot  | \hat\Theta_{k})$ and can be viewed as a population-level estimator, whereas $\hat\Theta_{k+1}$ in the sparse EM algorithm is obtained by maximizing the finite-sample $Q$-function $Q_y (\cdot | \hat\Theta_{k})$ and can be viewed as a perturbation to its population counterpart. Meanwhile, both $Q$-functions are conditioning on the sparse EM estimator $\hat\Theta_{k}$ from the previous iteration. We then break our theoretical analysis into two steps. We first characterize the contraction behavior of the intermediate estimator $\Theta_{k+1}$ at the population level. We next quantify the perturbation of the sparse EM estimator $\hat\Theta_{k+1}$ from $\Theta_{k+1}$. The resulting error bound of the sparse EM estimator consists of two errors, a computational error and a statistical error. The first comes from the population behavior, and the latter measures the perturbation due to finite samples. 

Our first step of theoretical analysis is to characterize the contraction behavior of $\Theta_{k+1}$. Define the oracle auxiliary function $q(\Theta) = Q (\Theta  |\Theta_*)$, and the maximizer $M(\Theta) = \argmax_{\Theta'} Q (\Theta' |\Theta)$. Also define a local neighborhood of the true parameters $\Theta_*$, for some constants $\lambda\in (0,1)$ and $r>0$, 
\begin{eqnarray} \label{eqn:B-neighbor}
\cB (\lambda , r) = \Big \{ \left\{\Ab, \sigma_{\eta}^2, \sigma_{\epsilon}^2 \right\} :   \   | \sigma_{\eta}^2 - \sigma_{\eta,*}^2| \le \lambda\sigma_{\eta,*}^2, \  |\sigma_{\epsilon}^2 - \sigma_{\epsilon,*}^2| \le \lambda \sigma_{\epsilon,*}^2 , \  \|\Ab - \Ab_* \|_{\max}\le r \Big \}. 
\end{eqnarray}
Our first key insight is that, in a local neighborhood of the true parameter $\Theta_*$, if $q(\cdot)$ is strongly concave, $Q(\cdot | \hat{\Theta}_k)$ is geometrically similar to $q(\cdot)$, and $M(\cdot)$ well behaves, then $\Theta_{k+1}$ is closer to the truth $\Theta_*$ than $\hat\Theta_k$. Denote the first and second-order partial derivatives of $q(\Theta)$ at any parameter entry $\theta$ in $\Theta$ as $\partial_\theta q(\Theta)$ and $\partial_\theta^2 q(\Theta)$. Then, for any $\Theta \in \cB(\lambda, r)$, $\partial_\theta^2 q(\Theta)$ is upper bounded by a negative constant. Therefore, $q(\Theta)$ is strongly concave in $\cB(\lambda, r)$. The next assumption characterizes the geometric similarity between $Q(\cdot | \hat{\Theta}_k)$ and $q(\cdot)$, as well as the behavior of $M(\cdot )$ in $\cB (\lambda, r)$. 
\begin{assumption} \label{asm: FOS}
For any  $\Theta \in \cB (\lambda , r)$ and any entry of the parameter $\theta$ in $\Theta$, assume that $|\partial_{\theta_M} q \{ M ( \Theta) \} | \le \kappa  |\partial_{\theta_*}^2 q (\Theta_*) | | \theta - \theta_*|$, for some constant $0<\kappa<1$, where $\theta_M$ and $\theta_*$ denote the corresponding parameter in $M(\Theta)$ and $\Theta_*$, respectively.
\end{assumption}

\noindent 
 We note that the above inequality always holds when $\Theta=\Theta_*$, even with $\kappa=0$. When $\kappa$ is strictly positive, intuitively, it is reasonable to extend this inequality over a local region $\cB(\lambda, r)$ around $\Theta_*$ with some positive $\lambda$ and $r$. A similar condition was also imposed in \cite{BWY17}. By Assumption \ref{asm: FOS}, the strong concavity of $q(\Theta)$, and the fact that $  \partial_{\theta*} q (\Theta_* ) = 0$, we have that, 
\begin{eqnarray*}
|\partial_{\theta_*}^2 q (\Theta_*) |  |  \theta_{k+1} - \theta_* |   \le  |\partial_{\theta_*} q (\Theta_*) -\partial_{\theta_{k+1}} q (\Theta_{k+1} ) |  \le \kappa |\partial_{\theta_*}^2 q (\Theta_*) | | \hat\theta_k - \theta_*|,
\end{eqnarray*}
for any corresponding entries $\theta_{k+1}, \hat\theta_k$ of $\Theta_{k+1}, \hat\Theta_k$. Therefore, the population update $\Theta_{k+1}$ is closer to $\Theta_*$ than $\hat{\Theta}_k$ at rate $\kappa$.

Next, we need to ensure the above contraction property holds for all population updates $\Theta_{k}$ for any $k \ge 1$. That is, once the initial estimator $\hat\Theta_0$ locates in $\cB(\lambda, r)$, then the population updates $\{\Theta_k\}_{k \ge 1}$ all locate in $\cB(\lambda, r)$, and thus the contraction property applies to any $\Theta_{k}$ with $k\ge 1$. To achieve that, we need to ensure that $M(\Theta) \in \cB(\lambda, r)$ for any $\Theta \in \cB(\lambda, r)$. Let $\bSigma_{y} (\Theta) \in \RR^{pT \times pT}$ be the covariance matrix of the stacked vector $\yb_{[T]}=(\yb_1^\top , \ldots , \yb_T^\top )^\top \in \RR^{pT}$ conditioning on the parameter set $\Theta$. 
Define $\bSigma_1 (\Theta) = (T-1)^{-1} \EE \left\{\sum_{t=1}^{T-1}   \EE ( \xb_{t}\xb_{t+1}^\top |  \yb_{[T]}, \Theta )  | \Theta_* \right\}$, $\bSigma_{0} (\Theta) = (T-1)^{-1} \EE \left\{\sum_{t=1}^{T-1}  \EE  ( \xb_{t}\xb_{t}^\top | \yb_{[T]}, \Theta )   | \Theta_* \right\}$, $\bSigma_{0}^{(1)} (\Theta) = (T-1)^{-1} \EE \left\{ \sum_{t=2}^{T}  \EE  ( \xb_{t}\xb_{t}^\top |  \yb_{[T]}, \Theta  )  | \Theta_* \right\}$. 
In a sense, the three terms can be viewed as expectations of $\Eb_{t, t +1, k}$ and $\Eb_{t, t , k}$ conditioning on true parameter if $\hat{\Theta}_{k-1}=\Theta$.

\begin{assumption}  \label{asm: concave}
For any $\Theta  = \{\Ab ,\sigma_{\eta}^2, \sigma_{\epsilon}^2\}  \in \cB(\lambda, r)$, assume that $\| \{\bSigma_1(\Theta)\}^\top \{ \bSigma_0(\Theta) \}^{-1}- \Ab_*\|_{\max} \le r$, $\Big | \tr \big [  \bSigma_{0}^{(1)} (\Theta) - \{ \bSigma_1 (\Theta) \}^\top \{\bSigma_0 (\Theta)\}^{-1} \bSigma_1(\Theta)  \big ]   - p {\sigma}_{\eta,*}^2 \Big | \le p \lambda \sigma_{\eta,*}^2$, and $\Big | pT   \sigma^2_{\epsilon}   +   \sigma^4_{\epsilon}\tr\big [  \{\bSigma_{y} (\Theta)\}^{-1 }  \bSigma_{y} (\Theta_*) \{\bSigma_{y} (\Theta)\}^{-1 }  - \{\bSigma_{y} (\Theta)\}^{-1 } \big ]  - pT {\sigma}^2_{\epsilon,*} \Big |  \le pT \lambda {\sigma}^2_{\epsilon,*}$.  
\end{assumption}

\noindent
This assumption trivially holds if $\Theta=\Theta_*$ when $\lambda=r=0$. When $\lambda, r >0$, intuitively, it is reasonable to expect the assumption remains valid over a proper local region around $\Theta_*$. 
Taking $\Ab_*$ for instance, note that $\{\bSigma_1(\hat{\Theta}_k)\}^\top \{ \bSigma_0(\hat{\Theta}_k) \}^{-1}$ is the population update of $\Ab_*$ from $\hat{\Theta}_k$ and falls in the region $\cB(\lambda, r)$ if $\hat{\Theta}_k \in \cB(\lambda, r)$, thus the identity $\bSigma_1(\Theta) = \bSigma_0(\Theta)\Ab^{\top}_* + (T-1)^{-1} \EE \left\{ \sum_{t=1}^{T-1}  \EE\left( \xb_{t}\bm{\eta}^{\top}_{t} | \{\yb_{t'}\}_{t'=1}^T, \Theta \right) \big | \; \Theta_* \right\}$ implies that this assumption essentially bounds the reminder $ (T-1)^{-1} \{\bSigma_0(\Theta)\}^{-1} \EE \left\{ \sum_{t=1}^{T-1}  \EE\left( \xb_{t}\bm{\eta}^{\top}_{t} | \{\yb_{t'}\}_{t'=1}^T, \Theta \right) \big | \; \Theta_* \right\}$, which should be small for a range of $\Theta_*$, since $\EE  ( \xb_{t}\bm{\eta}^{\top}_{t}| \Theta_*) = 0$. 
A similar condition was imposed in \citet[Condition C1]{CMZ19} on the initialization too. The explicit forms of $\lambda, r$ are difficult to obtain in our case though, given the temporal dependence and the complicated high-dimensional matrices and their inverses. 

Together, Assumptions  \ref{asm: FOS} and \ref{asm: concave} ensure that the population update $\Theta_{k+1}$ gets closer to the truth $\Theta_*$ than finite-sample update $\hat{\Theta}_k$ at a contraction rate $\kappa$. This subsequently leads to a geometrically decreasing computational error at  rate $\kappa$ in the error bound of the sparse EM estimator. 

Our second step of theoretical analysis is to quantify the perturbation of the sparse EM estimator $\hat\Theta_{k+1}$ from its population counterpart $\Theta_{k+1}$. We introduce the next two assumptions that characterize the temporal dependence and the sparsity of the model. In Assumption \ref{asm: stat_error_dep}, $ [ \Mb ]_{mn}$ denotes the $(m,n)$th $p\times p$ block matrix in $\Mb \in \RR^{Tp \times Tp}$.  

\begin{assumption}\label{asm: stat_error_dep}
Assume $\sup_{\Theta  \in \cB (\lambda , r)} \max_{m\in [T]} \sum_{l=1}^T \big\| \big[ \big\{ \bSigma_{y} (\Theta) \}^{-1} \bSigma_{y} (\Theta_*) \{\bSigma_{y} (\Theta) \big\}^{-1} \big]_{mn} \big\|_{\max}$ $<\infty$. In addition, assume that $\sup_{\Theta  \in \cB (\lambda , r)}\max_{m\in [T]} \sum_{n=1}^T \big( \big\| [\bSigma_{y} (\Theta_*) \{\bSigma_{y} (\Theta) \}^{-1 }]_{mn} \big\|_{\max} +$  $\big\|[\bSigma_{y} (\Theta_*) \{\bSigma_{y} (\Theta) \}^{-1}]_{mn} \big\|_{\max} \big) <\infty$.  
\end{assumption}

\noindent
This assumption constrains the temporal dependence, where the matrices in this assumption are the covariance matrices of the quadratic forms of $\yb_t$'s. Note that the randomness in $\hat{\Theta}_{k+1}$ comes from the average of the quadratic forms of $\yb_t$'s, and the law of large numbers holds as long as the temporal dependence between the quadratic forms of $\yb_t$'s is bounded. 
This assumption is reasonable as the inequalities that bound the spectral of the quadratic forms comply with the existing concentration theory \citep{NW11}. 

Next, consider a weakly sparse matrix space, $\cM(r_q , R_q, r_1,R_1)$, defined as, 
\begin{eqnarray} \label{eqn:M-space}
\left\{ \Mb \in \RR^{p\times p}: \max_{j\in [p]} \sum_{i=1}^p |M_{ij}|^q \le r_q ,   \sum_{i,j \in [p]} | M_{ij} |^q \le R_q,  \|\Mb\|_{l_1} \le r_1, \|\Mb\|_{1} \le R_1 \right\},
\end{eqnarray} 
for some constants $ 0\le q  <1$, $r_q> 0$, $R_q> 0$, $r_1> 0$, and $R_1> 0$.

\begin{assumption} \label{asm: stat_error_weaksparse}
There exist constants $q \in [0,1)$, $r_q> 0$, $R_q> 0$, and $r_1> 0$, such that, for any $\Theta \in \cB (\lambda, r)$, $\{ \bSigma_0 (\Theta) \}^{-1} \bSigma_1(\Theta)  \in \cM  ( r_q ,R_q,  r_1,R_1)$. 
\end{assumption}

\noindent
This assumption imposes a weak sparsity constraint on the matrix $\{ \bSigma_0 (\Theta) \}^{-1} \bSigma_1(\Theta)$, the population update of $\Ab_*$, in that the matrix can be dense as long as there are only a few dominant entries and the rest entries are small. Besides, we allow $ r_q, R_q$ in $\cM(r_q, R_q, r_1,R_1)$ to diverge in the subsequent theoretical development. This is much weaker than requiring the population update to be strictly sparse with only a few nonzero entries. This assumption is similar in spirit as the sparsity assumption in \citet{CMZ19}, except that it involves a more complicated form due to the temporal dependence of the time series model. 

Together, Assumptions \ref{asm: stat_error_dep} and \ref{asm: stat_error_weaksparse} ensure that the sparse EM estimator $\hat\Theta_{k+1}$ is not too far away from its population counterpart $\Theta_{k+1}$, which contributes to the statistical error in the error bound of the sparse EM estimator. 

Now we are ready to present the main theorem regarding the computational and statistical errors of our sparse EM estimator. The key idea is that, for any entry of the parameter space, we have that $|\theta_{k+1}- \theta_{*}| \le \kappa |\hat\theta_{k} - \theta_{*}|$. Besides, define $\delta_{\theta} = \sup_{\hat\Theta_k \in \cB(\lambda, r)} |\theta_{k+1} - \hat\theta_{k+1}|$ as the distance between the population and finite-sample estimator of $\theta$. Then, we have, 
\begin{eqnarray*}
|\hat{\theta}_{k+1} - \theta_{*}| \le |\theta_{k+1} - \theta_{*}| + |\hat{\theta}_{k+1} - \theta_{k+1}| 
\le \kappa |\hat\theta_{k}- \theta_{*}| + \delta_{\theta} 
\le \kappa^{k+1} |\hat{\theta}_{0}- \theta_{*}| + \frac{1}{1-\kappa}\delta_{\theta},
\end{eqnarray*}
in which the first term is the geometrically decaying computational error, and the second term is the statistical error. The next theorem gives a more precise summary. 

\begin{theorem} \label{thm: sem_consist}
Suppose the following conditions hold. 
\begin{enumerate}[(a)]
\item The initial parameter set $\hat\Theta_0 = \left\{ \hat{\Ab}_{0}, \hat{\sigma}_{\eta,0}^2, \hat{\sigma}_{\epsilon,0}^2 \right\}$ are in a neighborhood $\cB(\lambda, r )$ that satisfies Assumptions \ref{asm: FOS}, \ref{asm: concave}, \ref{asm: stat_error_dep}, and \ref{asm: stat_error_weaksparse} for some $\lambda \in (0,1)$ and $r>0$.
\item The tolerance parameter $\tau_l = c_l  (r_1+1) \sqrt{\log (p)/T}$ for some positive constant $c_l$, $\forall l \le k$. 
\item The dimension of time series $p$ and the length of series $T$ satisfy that $C \log p \le T$ for some positive constant $C$.
\end{enumerate}
Then, the sparse EM estimator $\hat\Theta_k = \left\{ \hat{\Ab}_{k}, \hat{\sigma}_{\eta,k}^2, \hat{\sigma}_{\epsilon,k}^2 \right\}$ at the $k$th iteration satisfies that, for any constant $c_0>0$, there exist positive constants $c_1$ to $c_5$ such that the event 
\begin{eqnarray*}
| \hat{\sigma}^2_{\epsilon,k} - \sigma^2_{\epsilon,*}| & \le & \kappa^{k} | \hat{\sigma}^2_{\epsilon,0} -  \sigma^2_{\epsilon, *} |  + \frac{c_1}{1-\kappa} \|\bSigma_{y } (\Theta_*)\|_2 \sup_{\Theta  \in \cB (\lambda, r)}\|\{\bSigma_{y } (\Theta )\}^{-1}\|_2^2  \sqrt{\frac{\log p}{ Tp}}, \\
| \hat{\sigma}^2_{\eta,k} -  \sigma^2_{\eta,*}| & \le & \kappa^{k} | \hat{\sigma}^2_{\eta,0 } -  \sigma^2_{\eta, *} |  +  \frac{c_2}{1-\kappa} \Bigg [ \| \bSigma_{y } (\Theta_*)\|_2 \left( 1 \vee \sup_{\Theta  \in \cB (\lambda, r)}\|\{\bSigma_{y } (\Theta )\}^{-1 }\|_2^2 \right) \sqrt{\frac{\log p}{T p}} \\ 
  & + &  \frac{1}{p}\sqrt{\frac{\log p}{T}} \Bigg\{ (r_1 \vee 1 ) R_1+ \; R_q \left[ (r_1 \vee 1 )\sqrt{\frac{\log p}{T}} \sup_{\Theta  \in \cB (\lambda, r)}  \big \|\{ \bSigma_0 (\Theta)\}^{-1} \big \|_{l_1} \right]^{1-q} \Bigg\} \Bigg], \\
\|\hat{\Ab}_{k} -\Ab_* \|_{\max} & \le & \kappa^{k} \|\hat{\Ab}_{0} -\Ab_* \|_{\max}  + \frac{c_3 }{1-\kappa} (r_1 \vee 1)\sup_{\Theta  \in \cB (\lambda, r)}  \big \|\{ \bSigma_0 (\Theta)\}^{-1} \big \|_{l_1} \sqrt{\frac{\log p}{T}}, \\
\|\hat{\Ab}_{k} -\Ab_* \|_{l_\infty} & \le & \kappa^{k} \|\hat{\Ab}_{0} -\Ab_* \|_{l_\infty} + \frac{c_4   }{1-\kappa} r_q \left[ (r_1 \vee 1) \sup_{\Theta  \in \cB (\lambda, r)}  \big \|\{ \bSigma_0 (\Theta)\}^{-1} \big \|_{l_1}  \sqrt{\frac{\log p}{T}} \right]^{1-q}, \\
 \|\hat{\Ab}_{k} -\Ab_*  \|_{r,2} & \le & \kappa^{k} \|\hat{\Ab}_{ 0} -\Ab_*  \|_{r,2}  + \frac{c_5  }{1-\kappa} \sqrt{r_q} \left[ (r_1 \vee 1)\sup_{\Theta  \in \cB (\lambda, r)}  \big \|\{ \bSigma_0 (\Theta)\}^{-1} \big \|_{l_1}  \sqrt{\frac{\log p}{T}} \right]^{1-\frac{q}{2}},
\end{eqnarray*}
happens with probability at least $1-p^{-c_0}$.
\end{theorem}

\noindent
We make some remarks. First, the non-asymptotic error bound portrays the estimation error of sparse EM at each iteration, and reveals the interplay between the computational efficiency and the statistical rate of convergence. After a sufficient number of iterations, the computational error is to be dominated by the statistical error. Second, the statistical errors are all vanishing if $\log p$ scales with $T$, and they decay sufficiently fast in terms of $p$ and $T$ for subsequent statistical inference, even when $r_q$ and $R_q$ in $\cM(r_q, R_q, r_1,R_1)$ in   \eqref{eqn:M-space} diverge. Third, the statistical errors for $\Ab_*$ do not explicitly display the sparsity. This information is hidden in $\cM(r_q, R_q, r_1,R_1)$. Moreover, since the update of $\sigma^2_{\eta,*}$ involves the update of $\Ab_*$, the statistical error of $\sigma^2_{\eta,*}$ is more complicated than that of $\sigma^2_{\epsilon,*}$. Finally, we observe the phenomenon of ``blessing of dimensionality'', in that the statistical errors of $\sigma^2_{\epsilon,*}$ and $\sigma^2_{\eta,*}$ decrease when the dimension $p$ grows under a fixed sample size $T$. In general, we allow $p$ to diverge at an exponential rate of $T$ as both approach infinity.

\section{Test Statistics} 
\label{sec: test-statistics}

We next construct a Gaussian matrix as our test statistic for the transition matrix inference in our high-dimensional VAR with measurement error. Given model \eqref{eq: model_measure}, we observe a time series of $\yb_t$ that follows an autoregressive structure, $\yb_{t+1} =  \Ab_* \yb_{t} + \eb_{t}$, with the error term $\eb_{t} = - \Ab_* \bepsilon_{t}+  \bepsilon_{t+1} +   \bm{\eta}_{t}$. Then the lag-1 auto-covariance of the error $\eb_t$ is of the form, 
\begin{equation*} 
\bSigma_e = \Cov(\eb_{t},\eb_{t-1}) = -\sigma_{\epsilon,*}^2 \Ab_*.
\end{equation*}
This suggests that we can apply the covariance testing methods on $\bSigma_e$ to infer transition matrix $\Ab_*$. However, $\eb_t$ is not directly observed. 
Define generic estimators of $\Theta_*$ by $\left \{\hat{\Ab},\hat\sigma_{\epsilon}^2, \hat\sigma_{\eta}^2 \right\}$. 
We use them to reconstruct this error, and obtain the sample lag-1 auto-covariance estimator, 
\begin{eqnarray*} 
\hat\bSigma_e = \frac{1}{T-2} \sum_{t=2}^{T-1} \hat{\eb}_{t}\hat{\eb}_{t-1}^\top, \quad \textrm{ where } \;\; 
\hat{\eb}_{t} = \yb_{t+1}  - \hat{\Ab} \yb_{t} - \frac{1}{T-1}\sum_{t'=1}^{T-1} (\yb_{t'+1} - \hat{\Ab}\yb_{t'}). 
\end{eqnarray*}
This sample estimator $\hat\bSigma_e$, nevertheless, involves some bias due to the reconstruction of the error term, and also an inflated variance due to the temporal dependence of the time series data. We next explicitly quantify such bias and variance, by characterizing the non-asymptotic behavior of $\hat\bSigma_e$, which eventually leads to our Gaussian matrix test statistic. 

Denote the maximal row-wise $\ell_1$ estimation error as $\Delta_{1} = \|\Ab_* - \hat\Ab\|_{\ell_1}$, and the maximal row-wise Euclidean estimation error as $\Delta_2=\|\Ab_* -\hat\Ab\|_{r,2}$. The next proposition characterizes the non-asymptotic behavior of $\hat\bSigma_e$. 

\begin{proposition} \label{thm: res_test}
For any constant $c >0$, there exist positive constants $c_1$ to $c_3$, such that, when $T \ge c_1 \log p$, 
\begin{align*}
\PP \left\{ \left\| \hat\bSigma_e + \left( \sigma_{\eta,*}^2 + \sigma_{\epsilon,*}^2 \right)\hat\Ab - \frac{1}{T-2} \sum_{t=2}^{T-1} {\eb}_{ t} {\eb}_{ t-1}^\top  \right\|_{\max}   \le c_2 \left( \Delta_{1} s_r \sqrt{ \frac{\log p}{T} } + \Delta_2^2 +  \frac{\log p}{T} \right) \right\} \\
\ge 1- c_3 p^{-c},
\end{align*}
where $s_r = \max_{i \in [p]} |\{j: A_{*,ij}\neq 0\}|$ is the maximal row-wise sparsity of $\Ab_{*}$. 
\end{proposition}

\noindent
This proposition suggests using $\sqrt{T-2}\hat\bSigma_e$ to construct the Gaussian matrix test statistic, since $(T-2)^{-1/2} \sum_{t=2}^{T-1} \left( {\eb}_{ t} {\eb}_{ t-1}^\top-\EE {\eb}_{ t} {\eb}_{ t-1}^\top \right)$ converges to a zero-mean Gaussian matrix by the central limit theorem. The max norm error of the sparse EM estimator and the fact $\EE \eb_{t}\eb_{t-1}^\top  = -\sigma_{\epsilon,*}^2 \Ab_*$ further imply that the non-vanishing bias of $\sqrt{T-2}\hat\bSigma_e$ is $\sqrt{T-2}\{-(\sigma_{\eta,*}^2 + \sigma_{\epsilon,*}^2)\hat{\Ab} + \sigma_{\eta,*}^2 \Ab_*\}$, which can be estimated by $\sqrt{T-2}\{-(\hat\sigma_{\eta}^2 + \hat\sigma_{\epsilon}^2) \hat{\Ab} + \hat\sigma_{\eta}^2 \Ab_0\}$ under the null hypothesis. Meanwhile, after the bias correction and some direct calculation of the entry-wise variance of $(T-2)^{-1/2} \sum_{t=2}^{T-1} \left( {\eb}_{ t} {\eb}_{ t-1}^\top-\EE {\eb}_{ t} {\eb}_{ t-1}^\top \right)$, the entry-wise limit variance of $\sqrt{T-2}\hat\bSigma_e$ is, 
\begin{eqnarray*}
\sigma_{*,ij}^2 & = & \left( \sigma_{\epsilon,*}^2 +\sigma_{\eta,*}^2 \right)^2 + \sigma_{\epsilon,*}^4 A_{*,ij}^2 +  2 \sigma_{\epsilon,*}^4  A_{*,ii} A_{*,jj}  + \sigma_{\epsilon,*}^4\|\Ab_{*,i:}\|_2^2\|\Ab_{*,j:}\|_2^2 \\
& & + \left( \sigma_{\epsilon,*}^4+\sigma_{\epsilon,*}^2 \sigma_{\eta,*}^2 \right)\left( \|\Ab_{*,i:}\|_2^2 + \|\Ab_{*,j:}\|_2^2 \right), \quad i,j \in [p]. 
\end{eqnarray*}
Plugging in the estimators $\left\{ \hat{\Ab}, \hat{\sigma}_\epsilon^2, \hat{\sigma}_\eta^2 \right\}$ into the above equation, we obtain the corresponding estimator $\hat{\sigma}_{ij}^2$. We also comment that, one can use any generic estimators $\left\{ \hat{\Ab}, \hat{\sigma}_\epsilon^2, \hat{\sigma}_\eta^2 \right\}$ to estimate the bias and variance of $\sqrt{T-2}\hat\bSigma_e$. Later, we present the sufficient conditions on the estimation precision of the generic estimators, so to achieve the desired theoretical properties of inference. We then show that the estimators from our sparse EM algorithm satisfy those conditions. 

Now, we construct the Gaussian matrix test statistic $\Hb$, whose $(i,j)$th entry is,  
\begin{equation} \label{eq: test-statistic}
H_{ij} = \frac{ \sum_{t=2}^{T-1}  \{ \hat{e}_{ t,i}\hat{e}_{ t-1,j} + \left( \hat{\sigma}_{\eta}^2 +\hat{\sigma}_{\epsilon}^2 \right)  \hat{A}_{ij}  - \hat{\sigma}_\eta^2 A_{0,ij} \} }{\sqrt{T-2} \; \hat{\sigma}_{ij}}, \quad i,j \in [p]. 
\end{equation}
Denote the estimation errors, $\Delta_{\epsilon}  = | \hat{\sigma}_{\epsilon}^2-\sigma_{\epsilon,*}^2 |$, $\Delta_{\eta} = | \hat{\sigma}_{\eta}^2- \sigma_{\eta,*}^2 |$, and $\Delta_{\sigma} = \max_{i,j \in [p]} | \hat{\sigma}_{ij}^2 -\sigma_{*,ij}^2 |$. The next theorem provides the sufficient conditions to guarantee the asymptotic standard normality of $H_{ij}$ under the null hypothesis. 

\begin{theorem} \label{thm: res_correction_normal}
Suppose the following conditions hold. 
\begin{enumerate}[(a)]
\item The estimation errors satisfy that $\Delta_{1} = o_p\left\{ s_r^{-1} (\log p)^{-1/2} \right\}, \Delta_2 = o_p (T^{-1/4}), \Delta_{\epsilon} = o_p(T^{-1/2}), \Delta_{\eta} = o_p(T^{-1/2})$, and $\Delta_{\sigma} = o_p (1)$. 
\item The dimension of time series $p$ and the length of series $T$ satisfy that $\log p = o(T^{1/2})$. 
\end{enumerate}
Then 
$$\frac{ \sum_{t=2}^{T-1}  \{\hat{e}_{ t,i}\hat{e}_{ t-1,j} + \left( \hat{\sigma}_{\eta}^2 +\hat{\sigma}_{\epsilon}^2 \right)  \hat{A}_{ij}  - \hat{\sigma}_\eta^2 A_{*,ij} \}}{\sqrt{T-2} \; \hat{\sigma}_{ij}}\convin{d} \N(0, 1)$$
uniformly for $i,j \in [p]$ as $p, T \to \infty$. 
\end{theorem}

\noindent 
Here the normality holds when the dimension $p$ grows at the exponential rate of $\sqrt{T}$. The matrix $\Hb$ is to serve as the test statistic for the subsequent inference procedures.

\section{Transition Matrix Inference}
\label{sec: inference}

\subsection{Global inference}

We first develop a testing procedure for the global hypotheses \eqref{eq: global_hypo}. The key observation is that the squared maximum entry of a zero mean normal vector converges to a Gumbel distribution \citep{CJ11}. Specifically, we construct the global test statistic as, 
\begin{equation*}
G_{\cS} = \max_{(i,j) \in \cS} H_{ij}^2. 
\end{equation*}
The next theorem states that the asymptotic null distribution of $G_{\cS}$ is Gumbel. We again state the sufficient conditions required for generic estimators $\left\{ \hat{\Ab}, \hat{\sigma}_\epsilon^2, \hat{\sigma}_\eta^2 \right\}$ first, and  show later that the sparse EM estimators satisfy these conditions.  

\begin{theorem}\label{thm: global_null}
Suppose the following conditions hold. 
\begin{enumerate}[(a)]
\item The estimation errors satisfy that $\Delta_{1} = o_p \left\{ (s_r\log p)^{-1} \right\}, \Delta_2 = o_p\left\{ (T\log p)^{-1/4} \right\}, \Delta_{\epsilon} = o_p\left\{ (T \log p)^{-1/2} \right\}, \Delta_{\eta} = o_p\left\{ (T \log p)^{-1/2} \right\}$, and $\Delta_\sigma = o_p\left\{ (\log p)^{-1} \right\}$.
\item The dimension of time series $p$ and the length of series $T$ satisfy that $\log p = o\left( T^{1/7} \right)$. 
\end{enumerate}
Then, under the global null hypothesis in \eqref{eq: global_hypo}, for any $\cS \subseteq [p] \times [p]$ and any $x \in \RR$, 
\begin{eqnarray*}
\lim_{|\cS| \rightarrow \infty} \PP \Big( G_\cS  -2 \log |\cS| + \log \log |\cS| \le x \Big) = \exp \left\{- \exp (-x/2) / \sqrt{\pi} \right\}.
\end{eqnarray*}
\end{theorem}

\noindent
We note that the condition (a) about the estimation consistency in Theorem \ref{thm: global_null} is stronger than that in Theorem \ref{thm: res_correction_normal} for the asymptotic normality. This is because the Gumbel convergence is built upon the normality property that needs to be guaranteed first. Based on this limiting null distribution, we define the asymptotic $\alpha$-level test as,
\begin{eqnarray*}
\Psi_\alpha = \ind \big[ G_\cS > 2 \log |\cS| -  \log \log |\cS| - \log \pi -2 \log\{-\log(1-\alpha)\} \big]. 
\end{eqnarray*}
We reject the global null if $\Psi_\alpha=1$. 

Next, we study the asymptotic power  of the test $\Psi_\alpha$. Toward that end, we introduce a parameter class of alternatives, 
\begin{equation} \label{eq: alter_class}
\cA (c, \cS) = \left\{ \left\{ \Ab_*, \sigma_{\eta,*}^2, \sigma_{\epsilon,*}^2 \right\}  : \max_{(i,j) \in \cS } \frac{\sigma_{\eta,*}^2 \delta_{ij}}{ \sigma_{*,ij}} \ge c \sqrt{\frac{\log |\cS|}{T}} \right\}, 
\end{equation}
where $\delta_{{ij}}=|A_{*,ij}-A_{0,ij}|$ is the distance between the null and the true transition matrix. The class $\cA (c , \cS)$ requires at least one entry in $\cS$ has a proper signal-to-noise ratio against the null. Note that this is a very large class, because the imposed magnitude $\sqrt{\log |\cS| /T}$ is vanishing, and it only requires one entry to satisfy. The next theorem shows that $\Psi_\alpha$ has the power converging to one uniformly over $\cA (2\sqrt{2}, \cS)$. Together, Theorems \ref{thm: global_null} and \ref{thm: global_power} establish the asymptotic size and power, and thus the consistency of the global test $\Psi_\alpha$. 

\begin{theorem}\label{thm: global_power}
Suppose the same conditions in Theorem \ref{thm: global_null} hold. Then 
\begin{eqnarray*}
\inf_{ \{\Ab_*, \sigma_{\eta,*}^2, \sigma_{\epsilon,*}^2\} \in \cA ( 2\sqrt{2}, \cS) } \PP (\Psi_\alpha =1 ) \to 1, \;\; \textrm{ as } |\cS| \to \infty. 
\end{eqnarray*}
\end{theorem}

Next, we show that, when we employ the sparse EM estimators developed in Section \ref{sec: EM}, we can obtain the same desired results as in Theorems \ref{thm: global_null} and \ref{thm: global_power}. Recall the sparse EM estimators at iteration $k$ are denoted as $\left\{ \hat{\Ab}_{k},\hat{\sigma}_{\eta,k}^2,\hat{\sigma}_{\epsilon,k}^2 \right\}$. Plugging in these estimators yields the corresponding sparse EM estimator $\hat{\sigma}_{ij,k}^2$ of $\sigma_{*,ij}^2$. Denote the global test statistic and the $\alpha$-level test based on these sparse EM estimators as $G_{\cS,\sEM}$ and $\Psi_{\alpha,\sEM}$, respectively. The next proposition establishes their size and power properties. 

\begin{proposition}\label{prop: EM_global}
Suppose the following conditions hold. 
\begin{enumerate}[(a)]
\item The initial parameter set $\hat\Theta_0 = \left\{ \hat{\Ab}_{0}, \hat{\sigma}_{\eta,0}^2, \hat{\sigma}_{\epsilon,0}^2 \right\}$ are in a neighborhood $\cB(\lambda, r )$ that satisfies Assumptions \ref{asm: FOS}, \ref{asm: concave}, \ref{asm: stat_error_dep}, and \ref{asm: stat_error_weaksparse} for some $\lambda \in (0,1)$ and $r>0$.

\item The parameters in \eqref{eqn:M-space} satisfy that $R_1 <\infty$, $r_q = o[ T^{\frac{1}{2}-\frac{q}{2}} / \{ s_r (\log p)^{\frac{3}{2}-\frac{q}{2}} \} ]$ and $R_q = o\{  p T^{\frac{1}{2}-\frac{q}{2}} /  (\log p)^{\frac{3}{2}-\frac{q}{2}} \}$. Moreover, $\sup_{\Theta \in \cB (\lambda, r)}\| \{\bSigma_{y} (\Theta) \}^{-1 }\|_2<\infty$, and $\|\bSigma_{y} (\Theta_*)\|_2 <\infty.$ 

\item The tolerance parameter $\tau_l = c_l  (r_1+1) \sqrt{\log (p)/T}$ for some positive constant $c_l$, $\forall l \le k$. 

\item The iteration $k\ge \lceil C \log\left\{ (T\log p) \vee (s_r \log p) \right\} \rceil$ for some positive constant $C$.

\item  The dimension of time series $p$ and the length of series $T$  satisfy that $\log p = o(T^{1/7})$.
\end{enumerate}
Then, under the global null hypothesis in \eqref{eq: global_hypo}, for any $\cS \subseteq [p] \times [p]$,
\begin{align*}
\lim_{|\cS| \rightarrow \infty} \PP \Big( G_{\cS,\sEM}  -2 \log |\cS| + \log \log |\cS| \le x \Big) & = \exp \left\{ - \exp(-x/2)/\sqrt{\pi} \right\}  \; \textrm{ for any } x \in \RR, \\
\inf_{ \{\Ab_*, \sigma_{\eta,*}^2, \sigma_{\epsilon,*}^2\} \in \cA ( 2\sqrt{2}, \cS) } \PP (\Psi_{\alpha,\sEM} =1 ) & \to 1, \; \textrm{ as }  \ |\cS| \to \infty.
\end{align*}
\end{proposition}

\noindent
The conditions for this proposition essentially combine those of Theorems \ref{thm: sem_consist} and \ref{thm: global_null}. When the number of iterations $k$ is large enough, the statistical error is to dominate the error bound of the sparse EM estimators, and this bound decays sufficiently fast to ensure the properties of the global testing procedure. The requirements on the statistical error of sparse EM are reasonable, in that $r_q$ and $R_q$ are allowed to diverge at certain rates. 
Moreover, we consider a finite $R_1$ here for simplicity, though it is possible to relax the sufficient condition to let $R_1$ diverge too.

\subsection{Simultaneous inference with FDR control}
 
We next develop a testing procedure for the simultaneous hypotheses \eqref{eq: multi_hypo} with a proper FDR control. Let $\cH_0 = \{(i,j) : A_{*,ij}=A_{0,ij}, (i,j) \in \cS \}$ denote the set of true null hypotheses, and $\cH_1 = \{ (i,j) : (i,j)\in \cS , (i,j) \notin \cH_0\}$ denote the set of true alternatives. The test statistic $H_{ij}$ follows a standard normal distribution when $H_{0;ij}$ holds, and as such, we reject $H_{0;ij}$ if $|H_{ij}| > t$ for some thresholding value $t > 0$. Let $R_{\cS}(t) = \sum_{(i,j) \in \cS} \ind \{ |H_{ij}|> t\}$ denote the number of rejections at $t$. Then the false discovery proportion (FDP) and the false discovery rate (FDR) in our simultaneous testing problem are,
\begin{eqnarray*}
\textrm{FDP}_{\cS}(t)=\frac{\sum_{(i,j) \in \cH_0} \ind \{ |H_{ij}|> t\}}{R_{\cS}(t)\vee 1}, \;\; \textrm{ and  } \;\; 
\textrm{FDR}_{\cS}(t) = \EE \left\{ \textrm{FDP}_{\cS}(t) \right\}.
\end{eqnarray*}
An ideal choice of the threshold $t$ is to reject as many true positives as possible, while controlling the false discovery at the pre-specified level $\beta$. That is, we choose $\inf \{ t > 0 : \text{FDP}_{\cS} (t) \le \beta \}$ as the threshold. However, $\cH_0$ in $\text{FDP}_{\cS} (t)$ is unknown. Observing that $\PP ( |H_{ij}|> t ) \approx 2\{ 1- \Phi (t) \}$ by Theorem \ref{thm: res_correction_normal}, where $\Phi (\cdot)$ is the cumulative distribution function of a standard normal distribution, we estimate the false rejections $\sum_{(i,j) \in\cH_0} \ind \{ |H_{ij}|> t\}$ in $\text{FDP}_{\cS} (t)$ using $\{ 2- 2 \Phi(t) \} |\cS|$. Moreover, we restrict the search of $t$ to the range $\left(0, \sqrt{2\log |\cS|} \right]$, since $\PP\left( \hat{t} \text{ exists in } \left(0, \sqrt{2\log |\cS|}\right] \right) \to 1$ as we show later in the proof of Theorem \ref{thm: FDR}. We summarize our simultaneous testing procedure in Algorithm \ref{alg: fdr_control}. 

\begin{algorithm}[t!]
\caption{Simultaneous inference with FDR control.} 
\label{alg: fdr_control}
\begin{algorithmic}
\STATE 1. Calculate   $H_{ij}$ for all $(i,j) \in \cS$. 
\STATE 2. Compute the thresholding value,  
\begin{equation*} \label{eq: search_hatt}
\hat{t} = \inf \cbr{0< t \le  \sqrt{2 \log |\cS|} :  \  \frac{ \{2- 2 \Phi(t)\} |\cS|}{R_{\cS} (t) \vee 1} \le \beta}. 
\end{equation*}
\ \ \ \ If $\hat{t}$ does not exist, set $\hat{t} = \sqrt{2 \log |\cS|}$. 
\STATE 3. For all $(i,j) \in \cS$, reject $H_{0; ij}$ if $|H_{ij}| > \hat{t}$. 
\end{algorithmic}
\end{algorithm}

Next, we study the asymptotic FDR control of Algorithm \ref{alg: fdr_control}. We need two assumptions. 

\begin{assumption} \label{asm: strong_signal} 
There exist positive constants $u_1$ and $u_2$, such that 
$$ \bigg | \bigg \{(i,j) : \   (i,j) \in  \cH_1, \frac{{\sigma}_{\eta,*}^2 \delta_{ij} }{{\sigma}_{*,ij} } > (4+u_1) \sqrt{\frac{\log p}{T}} \bigg \} \bigg | \ge u_2 \sqrt{\log\log |\cS|}.$$
\end{assumption}

\noindent 
This assumption requires a reasonable number of alternatives in $\cS$. Intuitively, if the number of alternatives is too small, then $\sum_{(i,j) \in \cH_0} \ind \{ |H_{ij}|> t\} \approx R_{\cS}(t)$ for any $t$, and the resulting FDR is close to one regardless thresholding value. This assumption is rather mild, since the required number is logarithm of logarithm of $|\cS|$. \cite{LS14} showed that this assumption is nearly necessary in the sense that the FDR control for large-scale simultaneous testing would fail if the number of true alternatives is fixed. 

\begin{assumption} \label{asm: bounded_dependence}
For some constants $0<v<(1-\bar{\sigma}) / (1+\bar{\sigma})$, $\gamma>0$, and $u>0$, we have $
\left| \left\{ \{(i_1,j_1),(i_2, j_2)\} : \left| \tilde{\sigma}_{i_1j_1, i_2j_2} \right| > (\log |\cS|)^{-2-\gamma} ; (i_1,j_1)\neq (i_2,j_2); (i_1,j_1) , (i_2, j_2) \in\cH_0  \right\} \right| \le u$ $|\cS|^{1+v}$, where $\tilde{\sigma}_{i_1j_1, i_2j_2}$ is the limit covariance between $H_{i_1j _1}$ and $H_{i_2 j_2}$, for $(i_1,j_1)\neq (i_2, j_2) \in \cS$, and $\bar{\sigma} = \max_{(i_1,j_1) \neq (i_2,j_2);(i_1,j_1), (i_2,j_2) \in  \cH_0} | \tilde{\sigma}_{i_1j_1, i_2j_2} |$. 
\end{assumption}

\noindent
This assumption bounds the number of strongly correlated entries in the null hypotheses. The bound, $|\cS|^{1+v}$, is weak, since there are $|\cS|^2$ pairs in total and the majority of them are allowed to be strongly correlated. A similar assumption was adopted in \cite{XCC18} to ensure the FDR control consistency. The explicit expression of $\tilde{\sigma}_{i_1j_1, i_2j_2}$ is given in the proof of Theorem \ref{thm: FDR}.

The next theorem shows that the simultaneous testing procedure in Algorithm \ref{alg: fdr_control} controls both FDR and FDP. 
We again state the sufficient conditions required for any estimators $\left\{ \hat{\Ab}, \hat{\sigma}_\epsilon^2, \hat{\sigma}_\eta^2 \right\}$ first, then show that the sparse EM estimators satisfy these conditions.

\begin{theorem} \label{thm: FDR}
Suppose the following conditions hold. 
\begin{enumerate}[(a)]
\item Suppose Assumptions \ref{asm: strong_signal} and \ref{asm: bounded_dependence} hold. 
\item The estimation errors satisfy the precision requirements in (a) of Theorem \ref{thm: global_null}. 
\item Suppose $|\cH_0| \ge c_1 |\cS|$ for some positive constant $c_1$. 
\item The dimension of time series $p$ and the length of series  $T$ satisfy that $p \le T^{c_2}$ for some positive constant $c_2$. 
\end{enumerate}
Then, for simultaneous hypotheses \eqref{eq: multi_hypo}, for any $\cS \subseteq [p] \times [p]$, 
\begin{align*}
\lim_{|\cS| \to \infty} \frac{\text{FDR}_{\cS} (\, \hat{t} \; )}{\beta |\cH_0|/|\cS|} = 1, \quad \textrm{ and } \quad 
\frac{\text{FDP}_{\cS} (\, \hat{t} \; )}{\beta | \cH_0|/|\cS|}\convin{p} 1 \;\;  \textrm{ as } \; |\cS| \to \infty.
\end{align*}
\end{theorem}
  
\noindent
Compared to the global testing, the estimation consistency condition (b) is the same for the simultaneous testing. Meanwhile, the simultaneous testing places some additional requirements on the number of alternatives as in condition (c) and Assumption \ref{asm: strong_signal}, and the entry dependence as in Assumption \ref{asm: bounded_dependence}. In addition, the dimension $p$ grows at the polynomial rate of the sample size $T$, as in condition (d). These requirements are reasonable because, intuitively, the global testing only deals with the maximum entry, whereas the simultaneous testing tackles every individual entry. As such, the simultaneous testing relies more on the dependence structure among the entries, and needs a larger sample size than the global testing. Finally, the slight deflation $\beta |\cH_0|/|\cS|$ in the limiting FDR comes from substituting $|\cH_0| $ with $|\cS|$ in the false rejection approximation.  

Next, we show that, when we employ the sparse EM estimators developed in Section \ref{sec: EM}, we can obtain the same properties as in Theorem \ref{thm: FDR}. 

\begin{proposition}\label{prop: EM_multiple}
Suppose the following conditions hold. 
\begin{enumerate}[(a)]
\item Suppose Assumptions \ref{asm: strong_signal} and \ref{asm: bounded_dependence} hold.

\item Suppose the conditions (a) to (d) in Proposition \ref{prop: EM_global} hold. 

\item Suppose $| \cH_0| \ge c_1 |\cS|$ for some positive constant $c_1$. 

\item The dimension of time series $p$ and the length of series  $T$ satisfy that $p \le T^{c_2}$ for some positive constant $c_2$. 
\end{enumerate}
Then, for simultaneous hypotheses \eqref{eq: multi_hypo}, for any $\cS \subseteq [p] \times [p]$,  
\begin{align*}
\lim_{|\cS| \to \infty} \frac{\text{FDR}_{\cS} (\, \hat{t} \; )}{\beta |\cH_0|/|\cS|} = 1, \quad \textrm{ and } \quad 
\frac{\text{FDP}_{\cS} (\, \hat{t} \; )}{\beta | \cH_0|/|\cS|}\convin{p} 1 \;\;  \textrm{ as } \; |\cS| \to \infty.
\end{align*}
\end{proposition}

\noindent 
The conditions for this proposition essentially combine those of Theorems \ref{thm: sem_consist} and \ref{thm: FDR}. The requirement (b) on the sparse EM algorithm is the same as that for the global testing.

\section{Simulations}
\label{sec: simulations}

\subsection{Setup}

\begin{figure}[t!]
\centering
\includegraphics[scale=0.23]{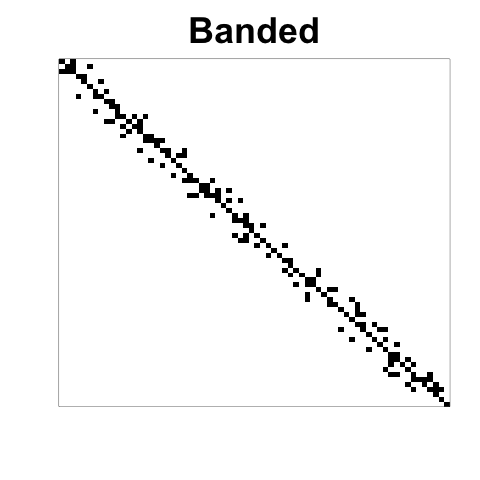}
\includegraphics[scale=0.23]{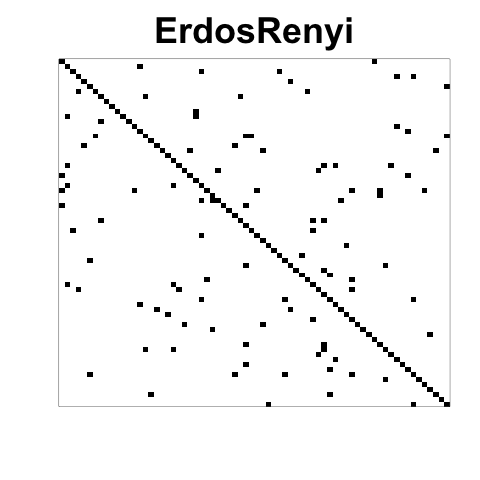}
\includegraphics[scale=0.23]{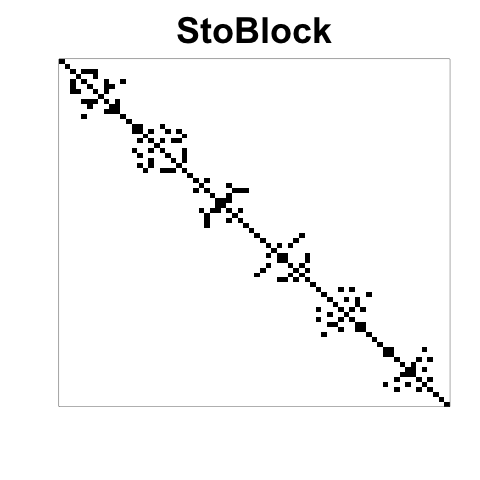}
\includegraphics[scale=0.23]{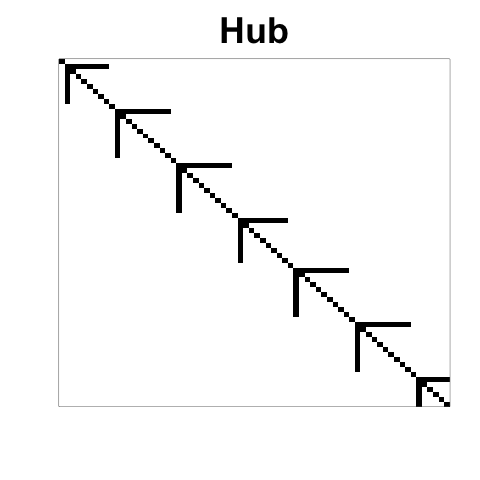}
\caption{Structures of the transition matrix $\Ab_*$. Black dots represent the nonzero entries. }
\label{fig: A_pattern}
\vskip 0em
\end{figure}

We carry out intensive simulations to study the finite-sample performance of our proposed method. We generate the data following model \eqref{eq: model_measure}. We consider four common network structures for the transition matrix $\Ab_*$: banded, Erd\"{o}s-R\'{e}nyi, stochastic block, and hub, as shown in Figure \ref{fig: A_pattern}. We first fix $\sigma_{\epsilon,*}=\sigma_{\eta,*}=0.2, \|\Ab_*\|_2=0.97$, and vary the dimension and sample size $(p,T)= (30,500), (50,500), (50,1000), (70,1000)$. Next, we fix $p=50, T=1000, \sigma_{\epsilon,*}=\sigma_{\eta,*}=0.2$, and vary the signal strength $\|\Ab_*\|_2= 0.7, 0.8, 0.9, 0.97$. Finally, we fix $p=50, T=1000, \|\Ab_*\|_2=0.97$, and vary the noise level $(\sigma_{\epsilon,*}, \sigma_{\eta,*})=(0.1,0.1), (0.2,0.2), (0.3,0.3), (0.4,0.4)$.  

\subsection{Parameter estimation}

We first report the estimation accuracy of our sparse EM. The tuning of the tolerance parameter $\tau_k$ in \eqref{eq: sparse_A} is done by cross-validation, where we use the first 25\% of data points for testing, the last 60\% for training, and the middle 15\% discarded to reduce the temporal dependence between the training and testing samples. 
We choose the value of $\tau_k$ that minimizes the average prediction error of the testing samples. We find our algorithm converges fast, usually within 10 iterations. 

\begin{figure}[t!]
\centering
\includegraphics[scale=0.17]{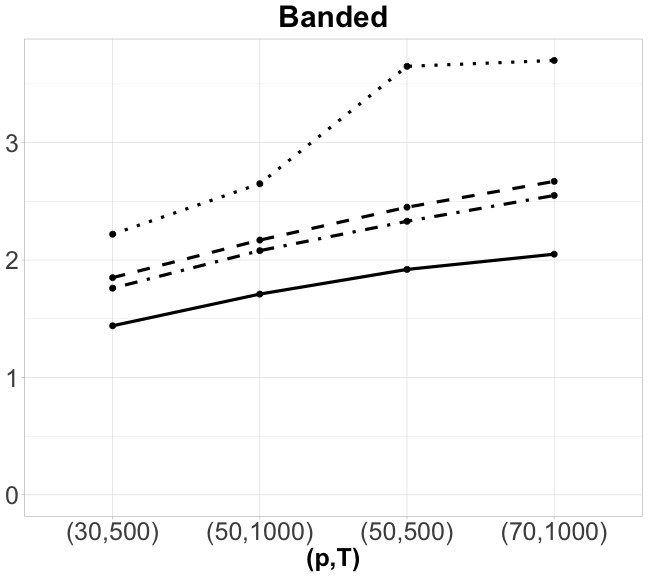}
\includegraphics[scale=0.17]{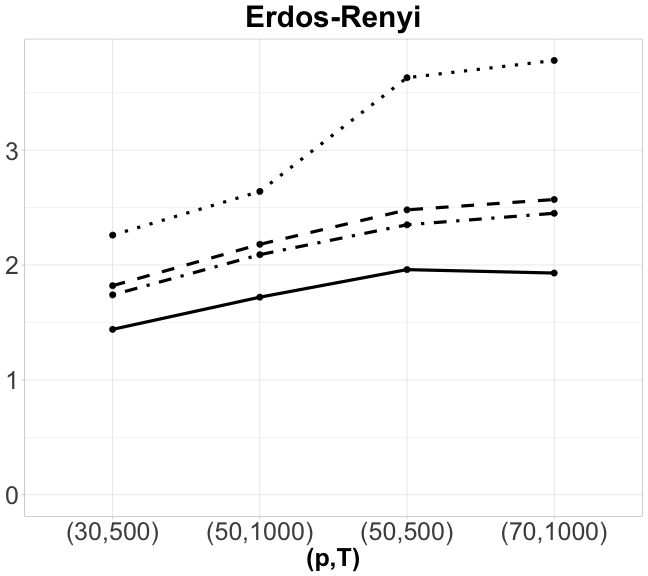}
\includegraphics[scale=0.17]{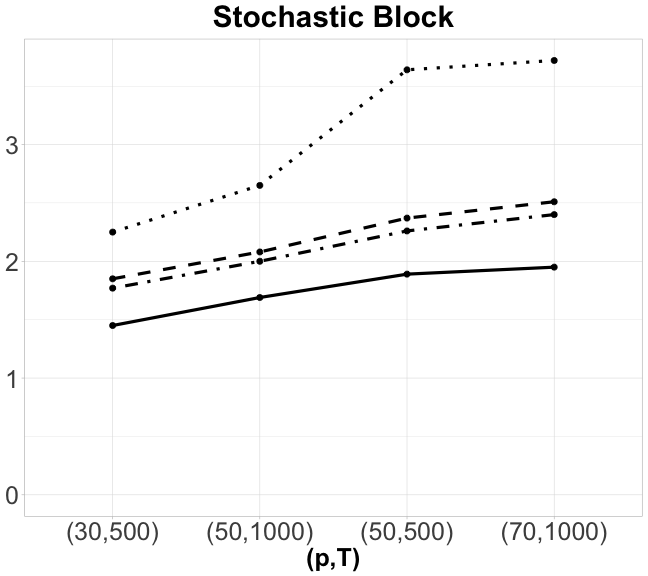}
\includegraphics[scale=0.17]{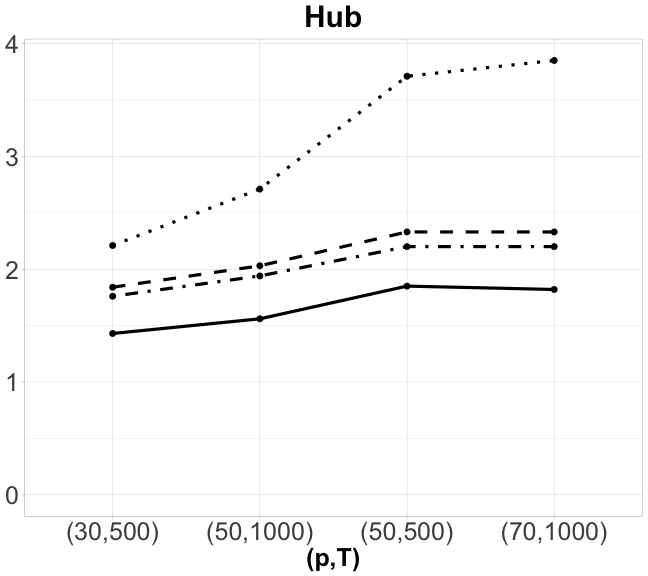}
\includegraphics[scale=0.17]{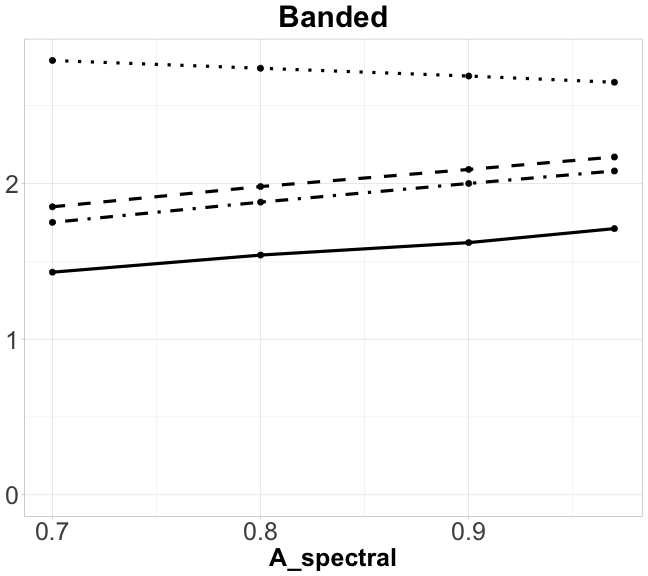}
\includegraphics[scale=0.17]{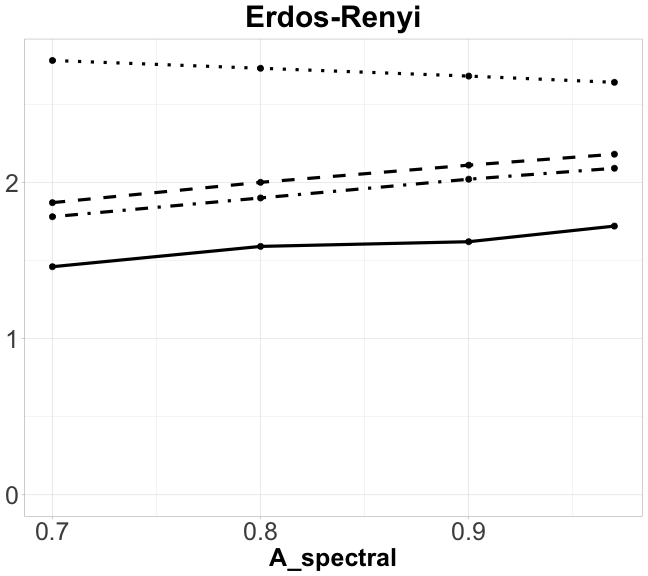}
\includegraphics[scale=0.17]{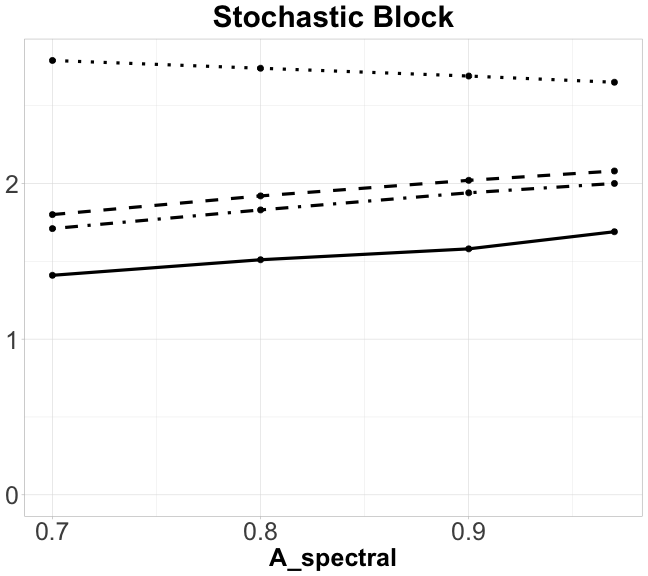}
\includegraphics[scale=0.17]{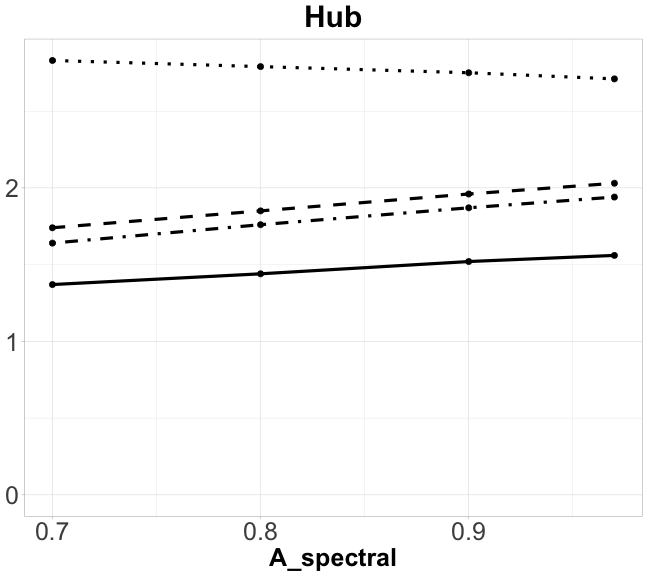}
\includegraphics[scale=0.17]{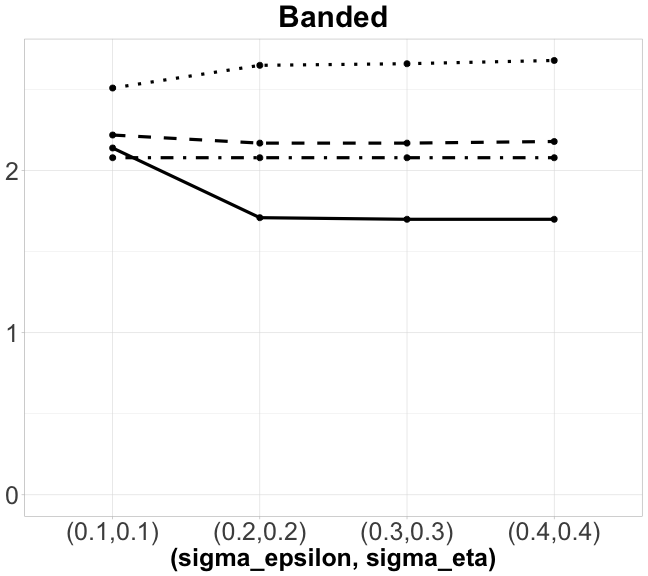}
\includegraphics[scale=0.17]{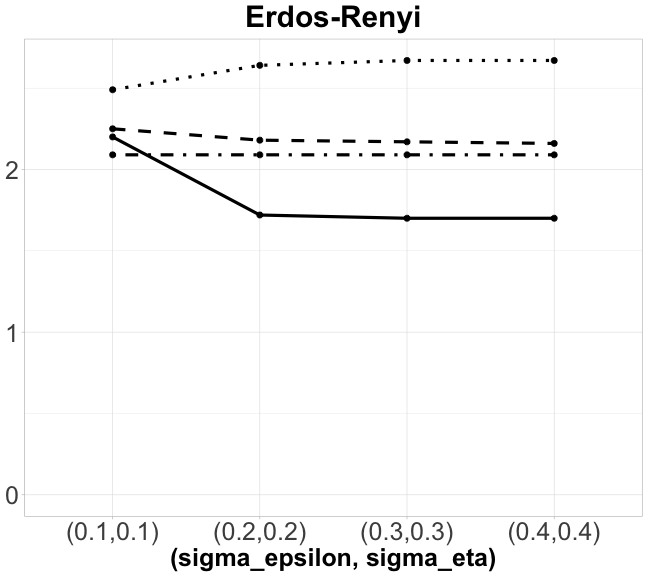}
\includegraphics[scale=0.17]{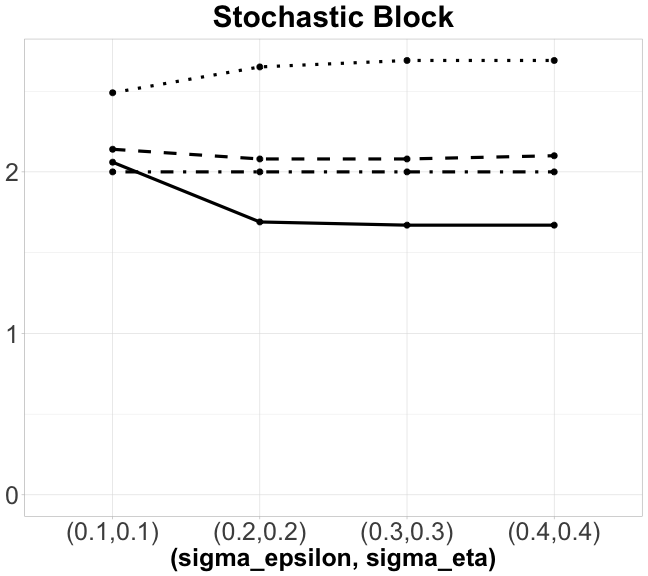}
\includegraphics[scale=0.17]{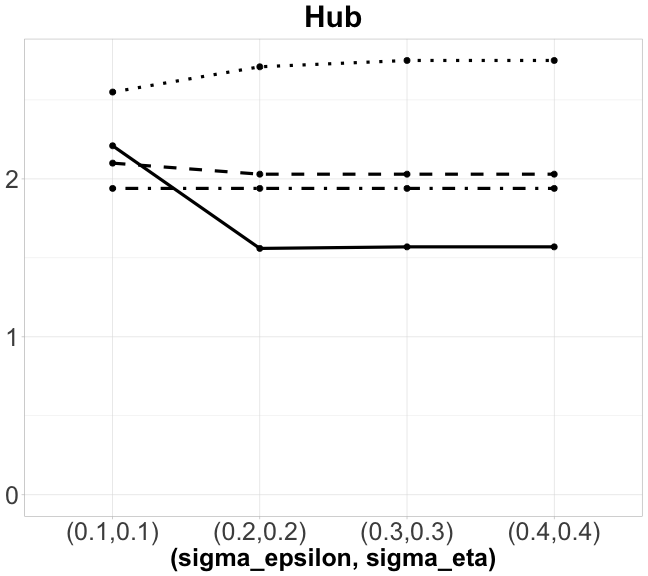}
\caption{Estimation error of the transition matrix $\Ab_*$ for four network structures, and the varying $(p, T)$ (top row), the varying signal strength $\|\Ab_*\|_2$ (middle row), and the varying noise level $(\sigma_{\epsilon,*}, \sigma_{\eta,*})$ (bottom row). Four methods are compared: the proposed sparse EM (solid line), the standard EM (dotted line), the Lasso estimator (dot-dashed line), and the Dantzig estimator (dashed line).}
\label{fig: A_frobenius}
\end{figure}

We compare our method with three  alternative solutions, including the standard EM without sparsity constraint, the Lasso estimator \citep{HHC08}, and the Dantzig estimator \citep{HLL15}, both of which were designed for VAR without measurement error. We evaluate the estimation accuracy by the Frobenius error $\nbr{\hat{\Ab} - \Ab_*}_F$. Figure \ref{fig: A_frobenius} reports the average estimation accuracy out of 1000 data replications for the varying $(p, T)$, the varying signal strength $\|\Ab_*\|_2$, and the varying noise level $(\sigma_{\epsilon,*}, \sigma_{\eta,*})$, respectively. It is seen that our proposed sparse EM achieves the smallest estimation error across all settings, except when the noise level is close to 0. For the case, the model reduces to a standard VAR model with little measurement error, and the Lasso and Dantzig estimators should work the best. Moreover, our method performs similarly under different network structures, reflecting its robustness with respect to the connectivity patterns. We also consider other error norms for $\Ab_*$ and the estimation accuracy for $\sigma_{\epsilon,*}, \sigma_{\eta,*}$. The results show the same qualitative patterns as Figure \ref{fig: A_frobenius}, and are thus omitted.

\subsection{Global and simultaneous inference}

We next evaluate the performance of our global and simultaneous inference procedures. Table \ref{tab: global_test} reports the empirical size and power based on 1000 data replications, with the significant level set at $\alpha =5\%$. It is seen that our global test maintains a reasonable control of the size, and at the same time achieves a good power. 
Table \ref{tab: multiple_test} reports the average false discovery proportion and the average true positive rate based on 1000 data replications, with the FDR level set at $5\%$. It is seen that our simultaneous test achieves both a high true positive rate and a low false discovery proportion.

\begin{table}[t!]
\centering 
\resizebox{\textwidth}{!}{
\begin{tabular}{c || c | c c || c | c c || c | c c} \hline\hline
&  $(p ,T)$ & Size & Power & $\|\Ab_*\|_2$ & Size & Power &  $(\sigma_{\epsilon,*}, \sigma_{\eta,*}) $ & Size & Power \\ \hline
banded & (30,500) & 3.1 & 100 & 0.7 & 3.4 & 100 & (0.1,0.1) & 4.5 & 100 \\ 
   &   & (0.17) & (0) &   & (0.18) & (0) &   & (0.21) & (0) \\ 
   & (50,500) & 2.7 & 100 & 0.8 & 3.1 & 100 & (0.2,0.2) & 2.4 & 100 \\ 
   &   & (0.16) & (0) &   & (0.17) & (0) &   & (0.15) & (0) \\ 
   & (50,1000) & 2.4 & 100 & 0.9 & 2.9 & 100 & (0.3,0.3) & 2.4 & 100 \\ 
   &   & (0.15) & (0) &   & (0.17) & (0) &   & (0.15) & (0) \\ 
   & (70,1000) & 2.7 & 100 & 0.97 & 2.4 & 100 & (0.4,0.4) & 2.4 & 100 \\ 
   &   & (0.16) & (0) &   & (0.15) & (0) &   & (0.15) & (0) \\ \hline 
Erd\"{o}s-R\'{e}nyi & (30,500) & 3.1 & 100 & 0.7 & 3.4 & 100 & (0.1,0.1) & 5.0 & 100 \\ 
   &   & (0.17) & (0) &   & (0.18) & (0) &   & (0.22) & (0) \\ 
   & (50,500) & 2.6 & 100 & 0.8 & 3.5 & 100 & (0.2,0.2) & 2.8 & 100 \\ 
   &   & (0.16) & (0) &   & (0.18) & (0) &   & (0.17) & (0) \\ 
   & (50,1000) & 2.8 & 100 & 0.9 & 3.1 & 100 & (0.3,0.3) & 2.7 & 100 \\ 
   &   & (0.17) & (0) &   & (0.17) & (0) &   & (0.16) & (0) \\ 
   & (70,1000) & 3 & 100 & 0.97 & 2.8 & 100 & (0.4,0.4) & 2.7 & 100 \\ 
   &   & (0.17) & (0) &   & (0.17) & (0) &   & (0.16) & (0) \\ 
   \hline    
stochastic block & (30,500) & 3.2 & 100 & 0.7 & 3.8 & 100 & (0.1,0.1) & 5.9 & 100 \\ 
   &   & (0.18) & (0) &   & (0.19) & (0) &   & (0.24) & (0) \\ 
   & (50,500) & 3.1 & 100 & 0.8 & 3.7 & 100 & (0.2,0.2) & 3.4 & 100 \\ 
   &   & (0.17) & (0) &   & (0.19) & (0) &   & (0.18) & (0) \\ 
   & (50,1000) & 3.4 & 100 & 0.9 & 3.5 & 100 & (0.3,0.3) & 3.3 & 100 \\ 
   &   & (0.18) & (0) &   & (0.18) & (0) &   & (0.18) & (0) \\ 
   & (70,1000) & 2.1 & 100 & 0.97 & 3.4 & 100 & (0.4,0.4) & 3.3 & 100 \\ 
   &   & (0.14) & (0) &   & (0.18) & (0) &   & (0.18) & (0) \\ 
      \hline 
hub  & (30,500) & 3.5 & 100 & 0.7 & 2.9 & 100 & (0.1,0.1) & 6.3 & 100 \\ 
   &   & (0.18) & (0) &   & (0.17) & (0) &   & (0.24) & (0) \\ 
   & (50,500) & 2 & 100 & 0.8 & 2.9 & 100 & (0.2,0.2) & 2.6 & 100 \\ 
   &   & (0.14) & (0) &   & (0.17) & (0) &   & (0.16) & (0) \\ 
   & (50,1000) & 2.6 & 100 & 0.9 & 2.5 & 100 & (0.3,0.3) & 2.6 & 100 \\ 
   &   & (0.16) & (0) &   & (0.16) & (0) &   & (0.16) & (0) \\ 
   & (70,1000) & 3.7 & 100 & 0.97 & 2.6 & 100 & (0.4,0.4) & 2.5 & 100 \\ 
   &   & (0.19) & (0) &   & (0.16) & (0) &   & (0.16) & (0) \\  \hline\hline
\end{tabular}
}
\caption{Empirical size and power, in percentage, of the global test for four network structures, and the varying $(p, T)$ (left column), the varying signal strength $\|\Ab_*\|_2$ (middle column), and the varying noise level $(\sigma_{\epsilon,*}, \sigma_{\eta,*})$ (right column). The standard errors are reported in the parentheses.} 
\label{tab: global_test}
\end{table}

\begin{table}[t!]
\centering 
\resizebox{\textwidth}{!}{
\begin{tabular}{c || c | c c || c | c c || c | c c} \hline\hline
 &  $(p ,T)$ & FDR & TPR & $\|\Ab_*\|_2$ &   FDR & TPR &  $(\sigma_{\epsilon,*}, \sigma_{\eta,*}) $ &  FDR & TPR \\ \hline
 banded & (30,500) & 4.34 & 73.65 & 0.7 & 4.59 & 71.42 & (0.1,0.1) & 5.39 & 93.59 \\ 
   &   & (0.03) & (0.05) &   & (0.02) & (0.04) &   & (0.02) & (0.02) \\ 
   & (50,500) & 3.91 & 67.19 & 0.8 & 4.45 & 82.55 & (0.2,0.2) & 3.91 & 92.3 \\ 
   &   & (0.02) & (0.05) &   & (0.02) & (0.03) &   & (0.02) & (0.02) \\ 
   & (50,1000) & 3.91 & 92.3 & 0.9 & 4.18 & 89.24 & (0.3,0.3) & 3.91 & 92.27 \\ 
   &   & (0.02) & (0.02) &   & (0.02) & (0.03) &   & (0.02) & (0.02) \\ 
   & (70,1000) & 3.73 & 88.44 & 0.97 & 3.91 & 92.3 & (0.4,0.4) & 3.91 & 92.27 \\ 
   &   & (0.02) & (0.02) &   & (0.02) & (0.02) &   & (0.02) & (0.02) \\ 
           \hline
Erd\"{o}s-R\'{e}nyi & (30,500) & 3.95 & 75.45 & 0.7 & 4.63 & 70.53 & (0.1,0.1) & 5.61 & 98.27 \\ 
   &   & (0.03) & (0.06) &   & (0.02) & (0.05) &   & (0.02) & (0.01) \\ 
   & (50,500) & 3.93 & 65.68 & 0.8 & 4.59 & 86.83 & (0.2,0.2) & 3.98 & 97.4 \\ 
   &   & (0.02) & (0.05) &   & (0.02) & (0.03) &   & (0.02) & (0.02) \\ 
   & (50,1000) & 3.98 & 97.4 & 0.9 & 4.21 & 94.57 & (0.3,0.3) & 3.97 & 97.35 \\ 
   &   & (0.02) & (0.02) &   & (0.02) & (0.02) &   & (0.02) & (0.02) \\ 
   & (70,1000) & 4.07 & 91.36 & 0.97 & 3.98 & 97.4 & (0.4,0.4) & 3.97 & 97.34 \\ 
   &   & (0.02) & (0.02) &   & (0.02) & (0.02) &   & (0.02) & (0.02) \\ 
           \hline
stochastic block & (30,500) & 4.14 & 73.96 & 0.7 & 4.7 & 66.79 & (0.1,0.1) & 5.62 & 90.98 \\ 
   &   & (0.03) & (0.05) &   & (0.03) & (0.05) &   & (0.02) & (0.02) \\ 
   & (50,500) & 3.75 & 61.12 & 0.8 & 4.65 & 78.86 & (0.2,0.2) & 4.18 & 89.6 \\ 
   &   & (0.02) & (0.05) &   & (0.02) & (0.04) &   & (0.02) & (0.03) \\ 
   & (50,1000) & 4.18 & 89.6 & 0.9 & 4.42 & 86.15 & (0.3,0.3) & 4.17 & 89.55 \\ 
   &   & (0.02) & (0.03) &   & (0.02) & (0.03) &   & (0.02) & (0.03) \\ 
   & (70,1000) & 3.97 & 84.63 & 0.97 & 4.18 & 89.6 & (0.4,0.4) & 4.18 & 89.54 \\ 
   &   & (0.02) & (0.02) &   & (0.02) & (0.03) &   & (0.02) & (0.03) \\ 
      \hline
hub & (30,500) & 4.33 & 81.02 & 0.7 & 4.75 & 65.07 & (0.1,0.1) & 6.37 & 96.96 \\ 
   &   & (0.03) & (0.05) &   & (0.03) & (0.05) &   & (0.02) & (0.02) \\ 
   & (50,500) & 3.81 & 58.74 & 0.8 & 4.7 & 82.08 & (0.2,0.2) & 4.25 & 95.28 \\ 
   &   & (0.02) & (0.06) &   & (0.02) & (0.04) &   & (0.02) & (0.02) \\ 
   & (50,1000) & 4.25 & 95.28 & 0.9 & 4.45 & 91.76 & (0.3,0.3) & 4.26 & 95.28 \\ 
   &   & (0.02) & (0.02) &   & (0.02) & (0.03) &   & (0.02) & (0.02) \\ 
   & (70,1000) & 4.39 & 77.21 & 0.97 & 4.25 & 95.28 & (0.4,0.4) & 4.19 & 95.26 \\ 
   &   & (0.02) & (0.03) &   & (0.02) & (0.02) &   & (0.02) & (0.02) \\ \hline\hline
\end{tabular}
}
\caption{Average false discovery proportion (FDP) and true positive rate (TPR), in percentage, of the simultaneous test for four network structures, and the varying $(p, T)$ (left column), the varying signal strength $\|\Ab_*\|_2$ (middle column), and the varying noise level $(\sigma_{\epsilon,*}, \sigma_{\eta,*})$ (right column). The standard errors are reported in the parentheses.} 
\label{tab: multiple_test}
\end{table}

\section{Brain Connectivity Analysis} 
\label{sec: HCP}

We illustrate the proposed method with a brain connectivity study based on task-evoked fMRI. The data is part of the Human Connectome Project \citep[HCP,][]{van2013wu}, whose overarching objective is to understand brain connectivity patterns of healthy adults. We study the fMRI scans of two individual subjects of the same age and sex and both participating the same story-math task. The task consists of blocks of auditory stories and addition-subtraction calculations, and requires the participant to answer a series of questions. An accuracy score is given at the end based on the participant's answers. The performance of the two subjects differ considerably, with one achieving the perfect score and the other getting only about half correct. We aim to estimate and infer the brain connectivity networks of the two subjects and compare between them.  We have pre-processed the fMRI data following the pipeline of \citet{glasser2013minimal}. The resulting data for each subject are $p=264$ time series, corresponding to 264 brain regions-of-interest following the brain atlas of \cite{power2011functional}. The length of each time series is $T=316$. The 264 brain regions have been further grouped into 14 functional modules \citep{Smith2009}: auditory (AD), cerebellar (CR), cingulo-opercular task control (CO), default mode (DM), dorsal attention (DAT), fronto-parietal  task control (FP), memory retrieval (MR), salience (SA), sensory/somatomotor  hand (SMH), sensory/somatomotor mouth (SMM), subcortical (SC), uncertain (UN), ventral  attention (VA), and visual (VS). Each module possesses a relatively autonomous functionality, and complex brain tasks are believed to perform through coordinated collaborations among the modules.

\begin{figure}[t!]
\centering
\includegraphics[scale=0.45]{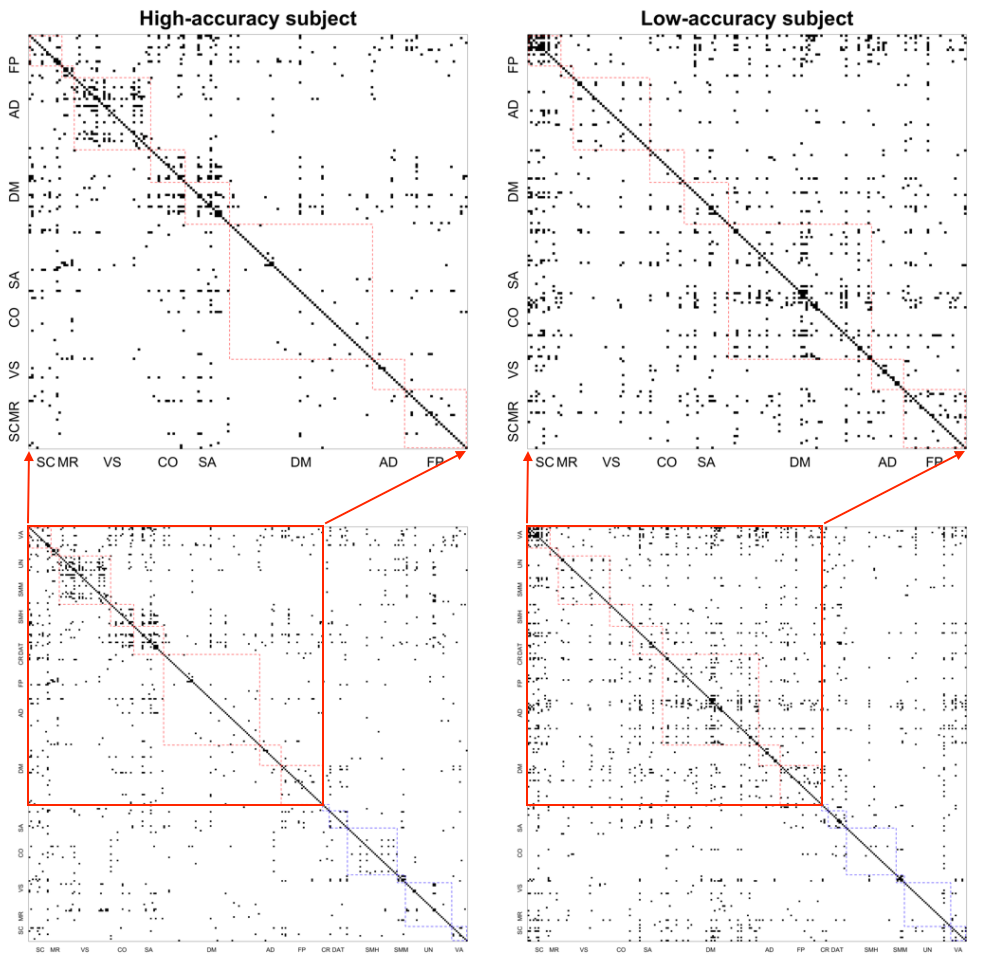}
\caption{The identified brain connectivity patterns for the high-accuracy subject (left column) and low-accuracy subject (right column). The 14 functional modules are indicated by the blocks (bottom row), and the 8 modules that demonstrate the most within-module connections are highlighted and amplified (top row).}
\label{fig: HCP_mat}
\vskip 0em
\end{figure}

\begin{figure}[h!]
\centering
\begin{tabular}{ccccc}
\multicolumn{2}{c}{\textbf{Subcortical}} & \hbox{     } & \multicolumn{2}{c}{\textbf{Memory retrieval}} \\
\includegraphics[scale=0.085]{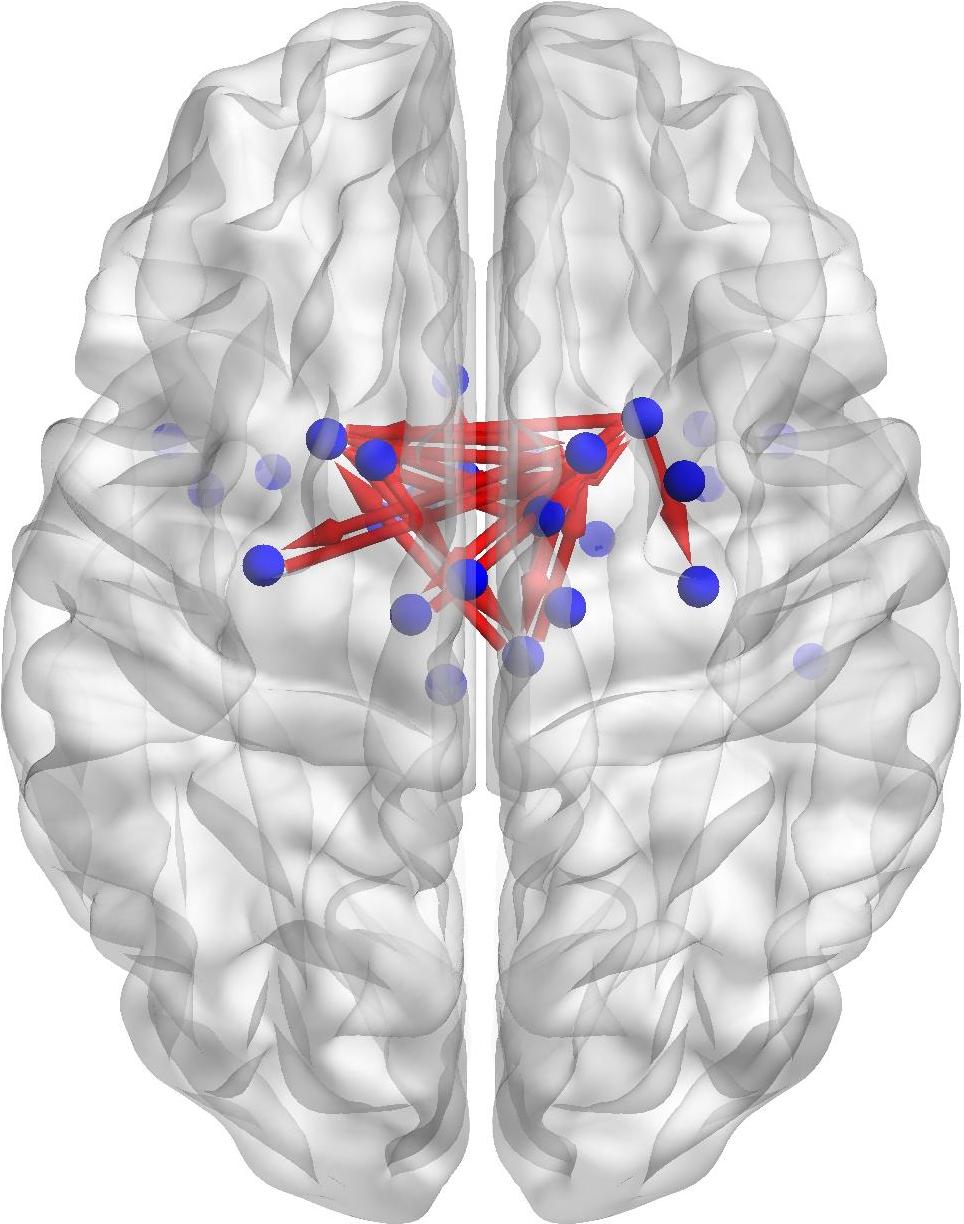} & 
\includegraphics[scale=0.085]{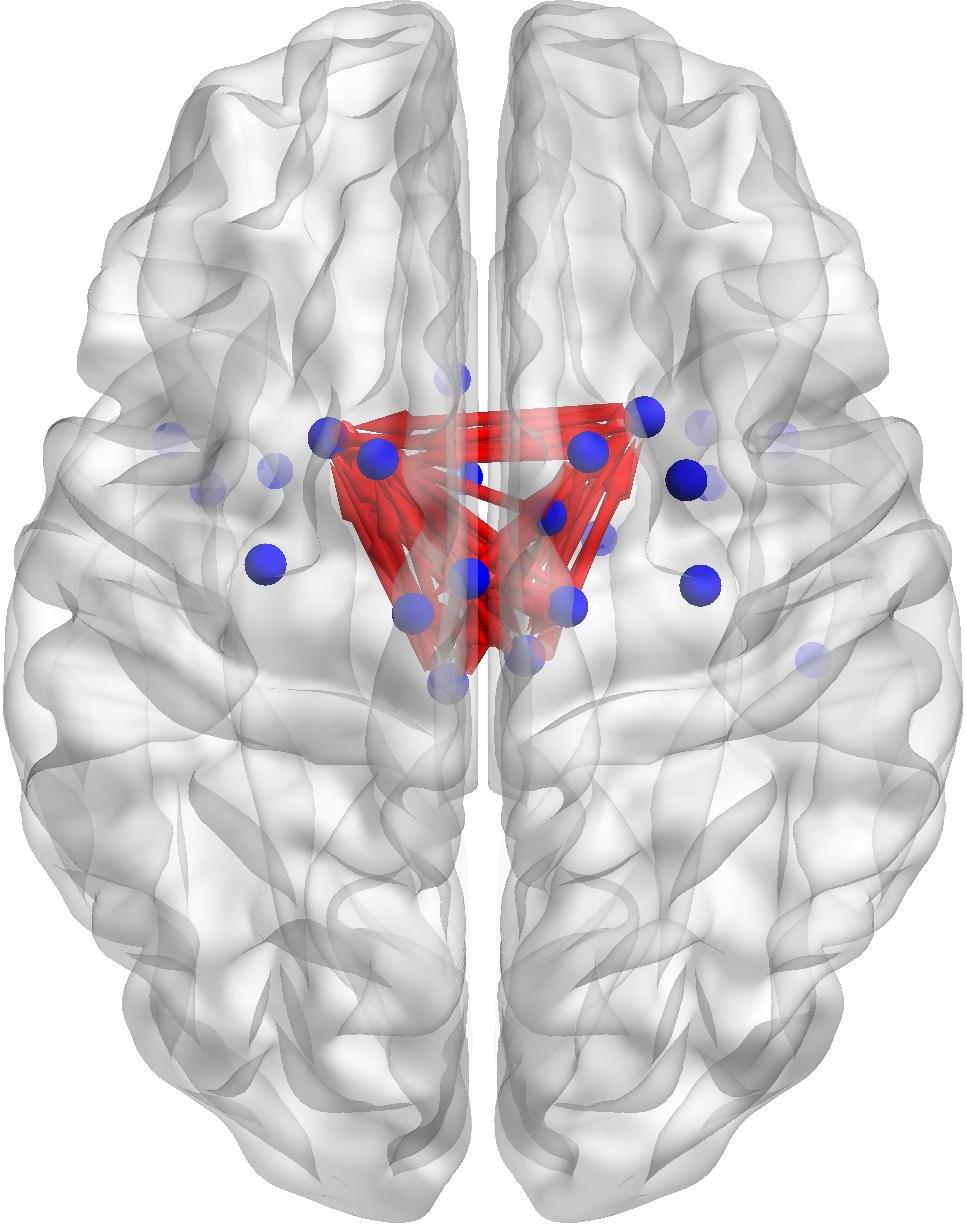} & 
&
\includegraphics[scale=0.085]{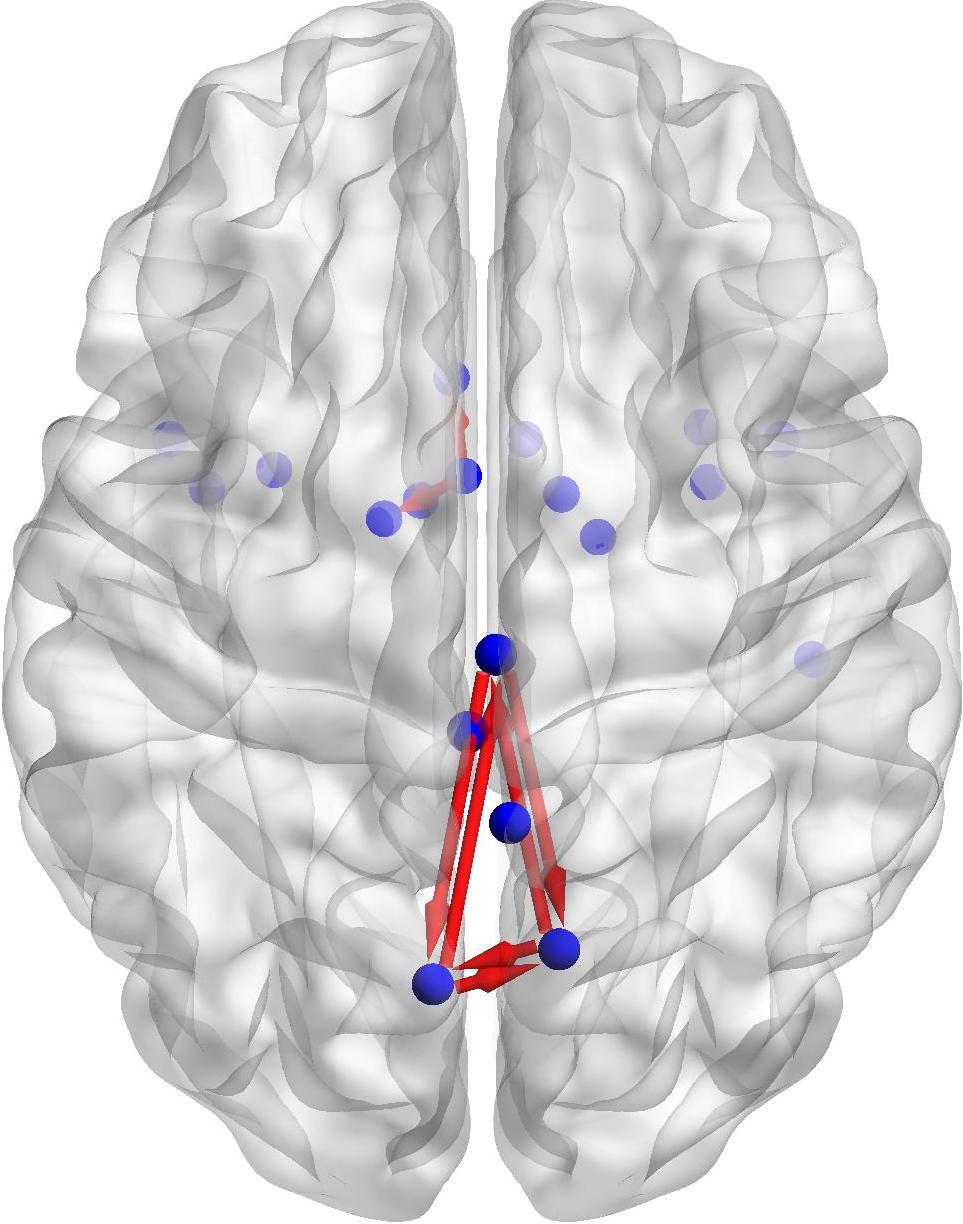} & 
\includegraphics[scale=0.085]{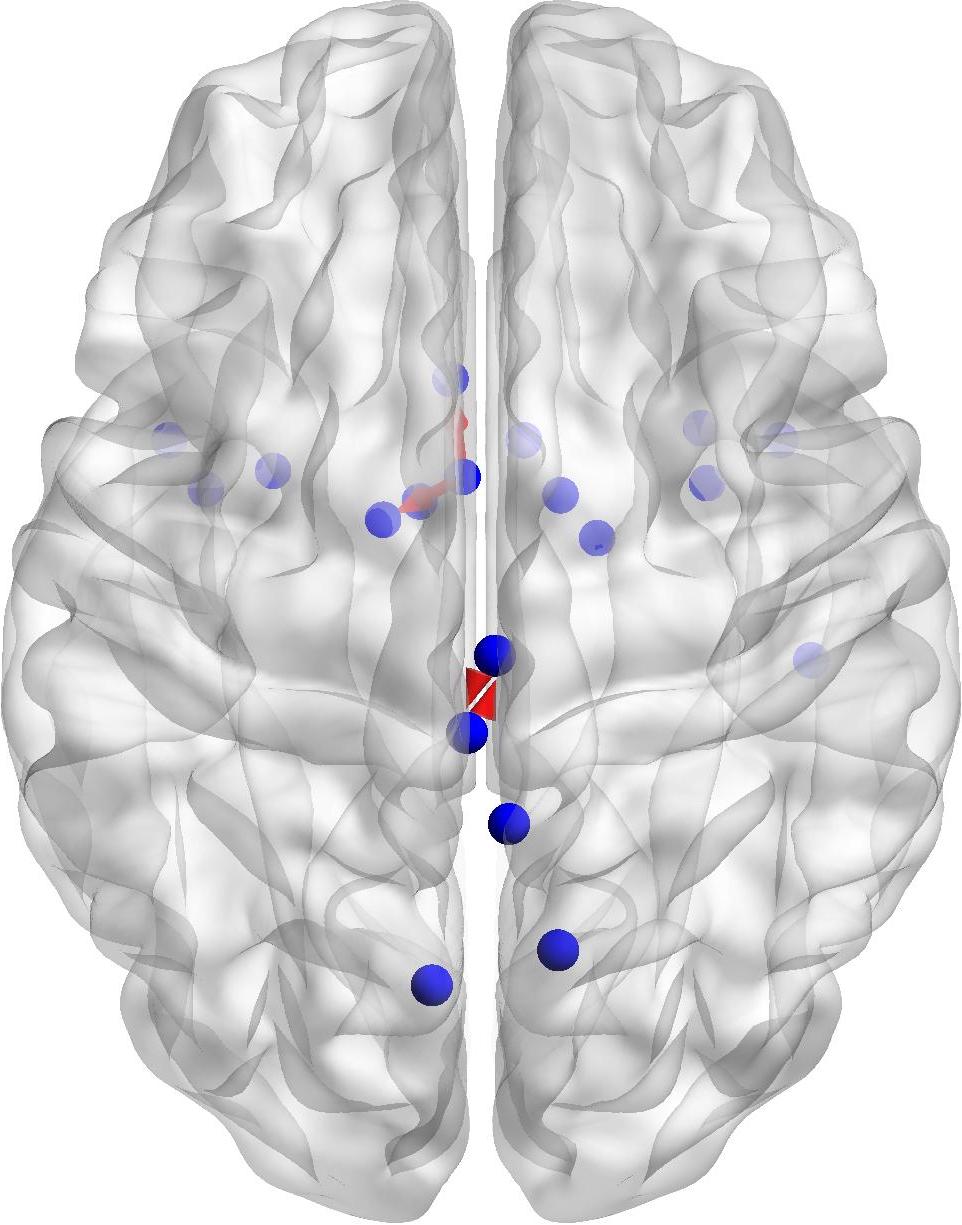} \\
high-accuracy & low-accuracy & & high-accuracy & low-accuracy \\
\\
\multicolumn{2}{c}{\textbf{Visual}} & \hbox{     } & \multicolumn{2}{c}{\textbf{Cingulo-opercular task control}} \\
\includegraphics[scale=0.085]{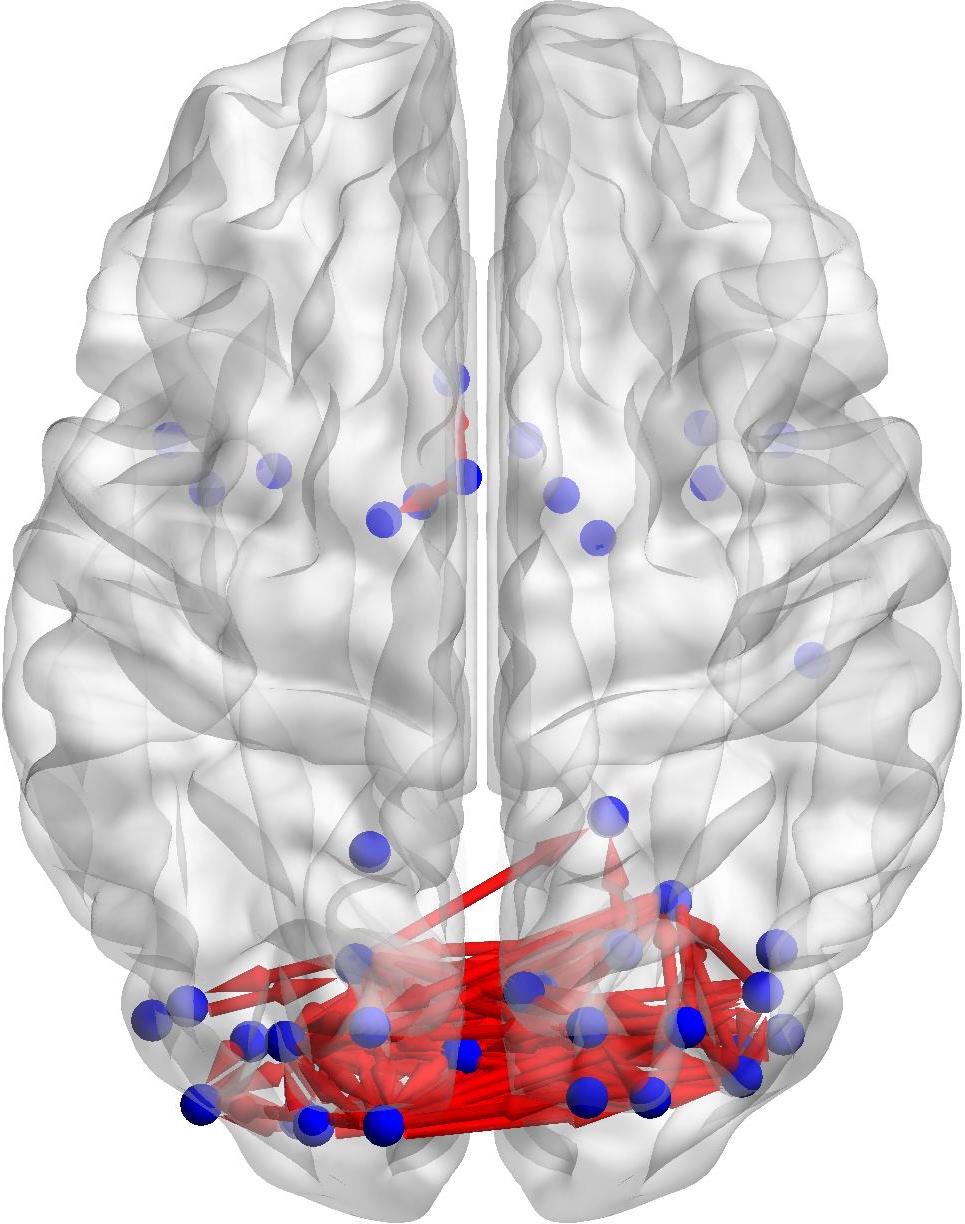} & 
\includegraphics[scale=0.085]{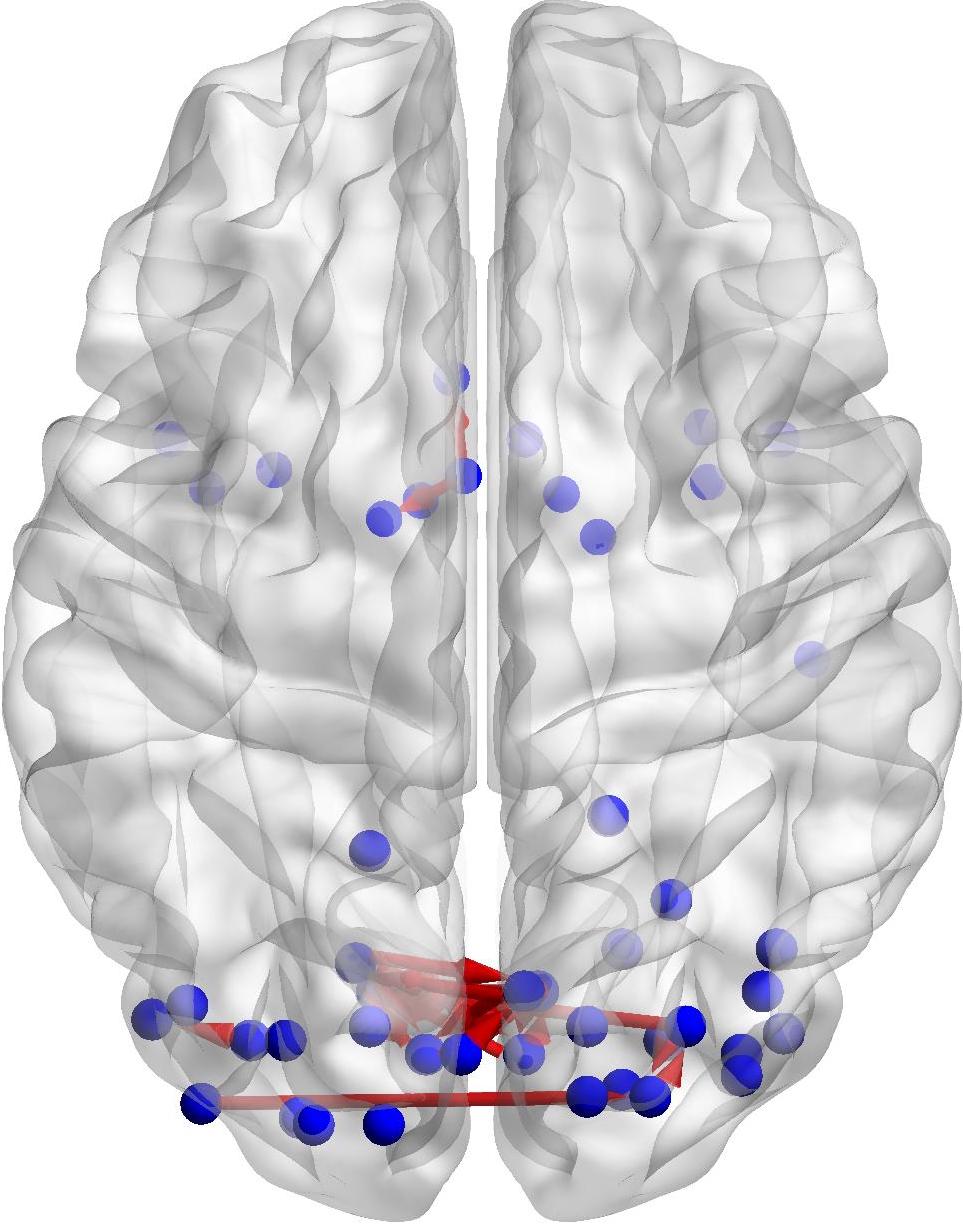} & 
&
\includegraphics[scale=0.085]{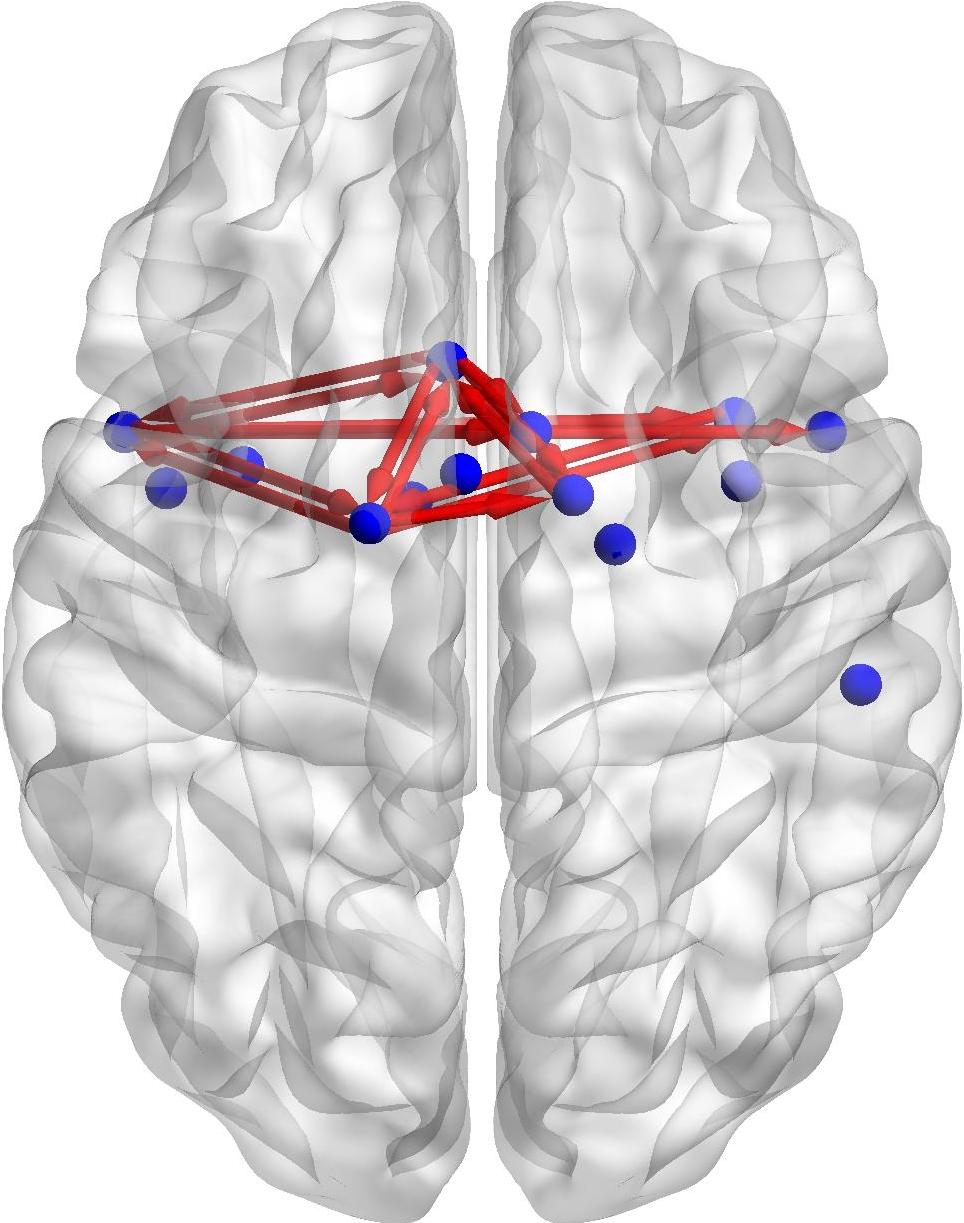} & 
\includegraphics[scale=0.085]{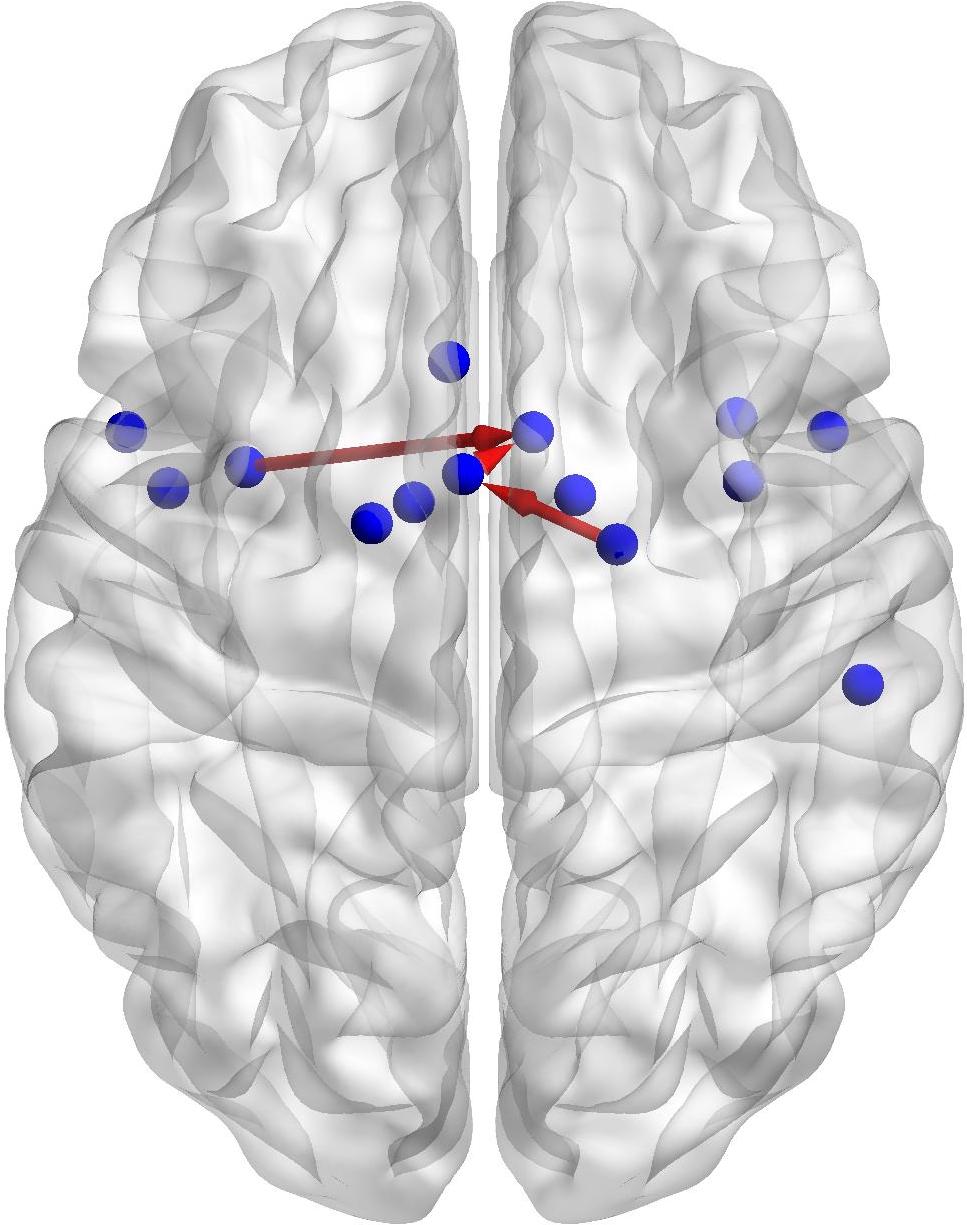} \\
high-accuracy & low-accuracy & & high-accuracy & low-accuracy \\
\\
\multicolumn{2}{c}{\textbf{Salience}} & \hbox{     } & \multicolumn{2}{c}{\textbf{Default mode}} \\
\includegraphics[scale=0.085]{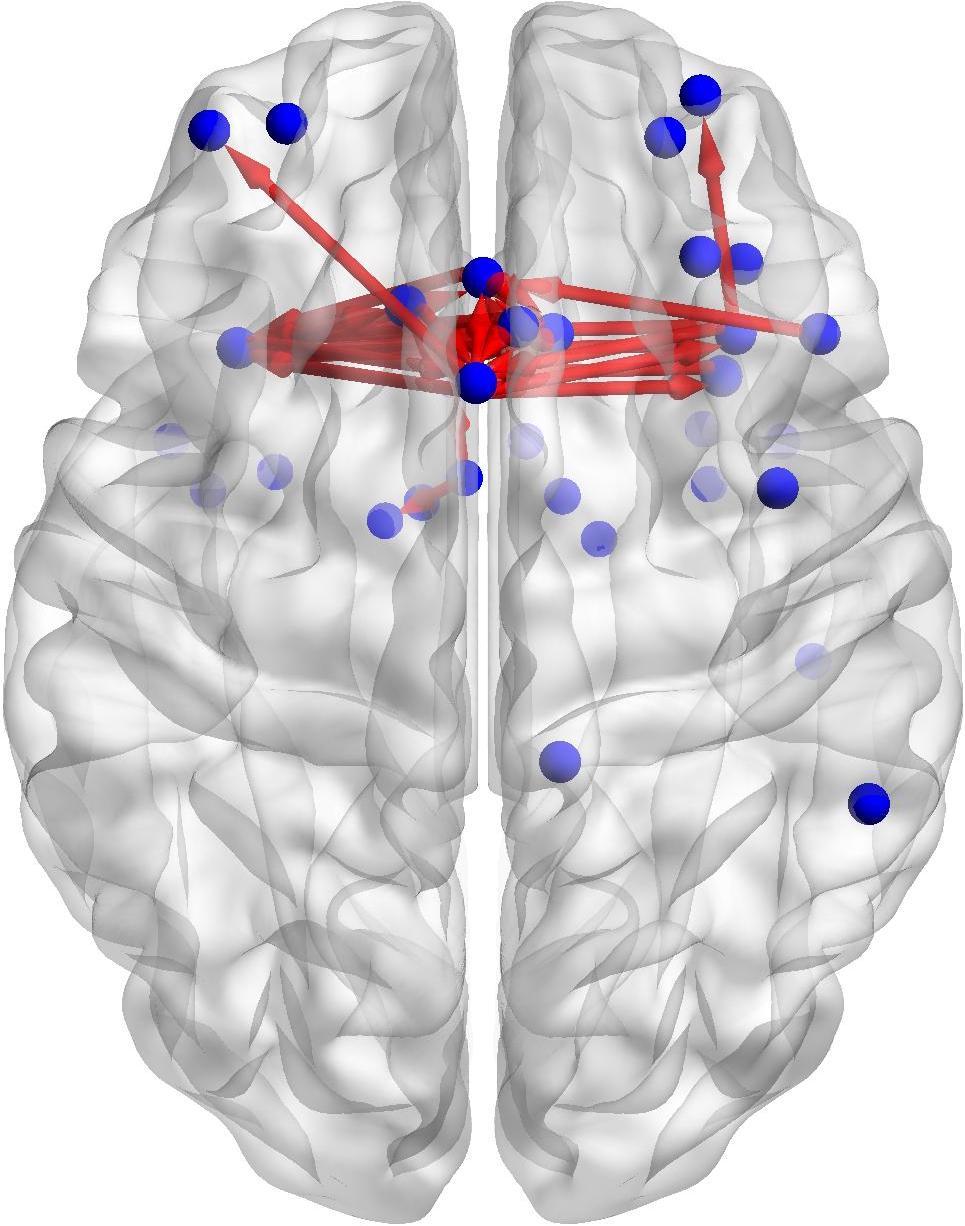} & 
\includegraphics[scale=0.085]{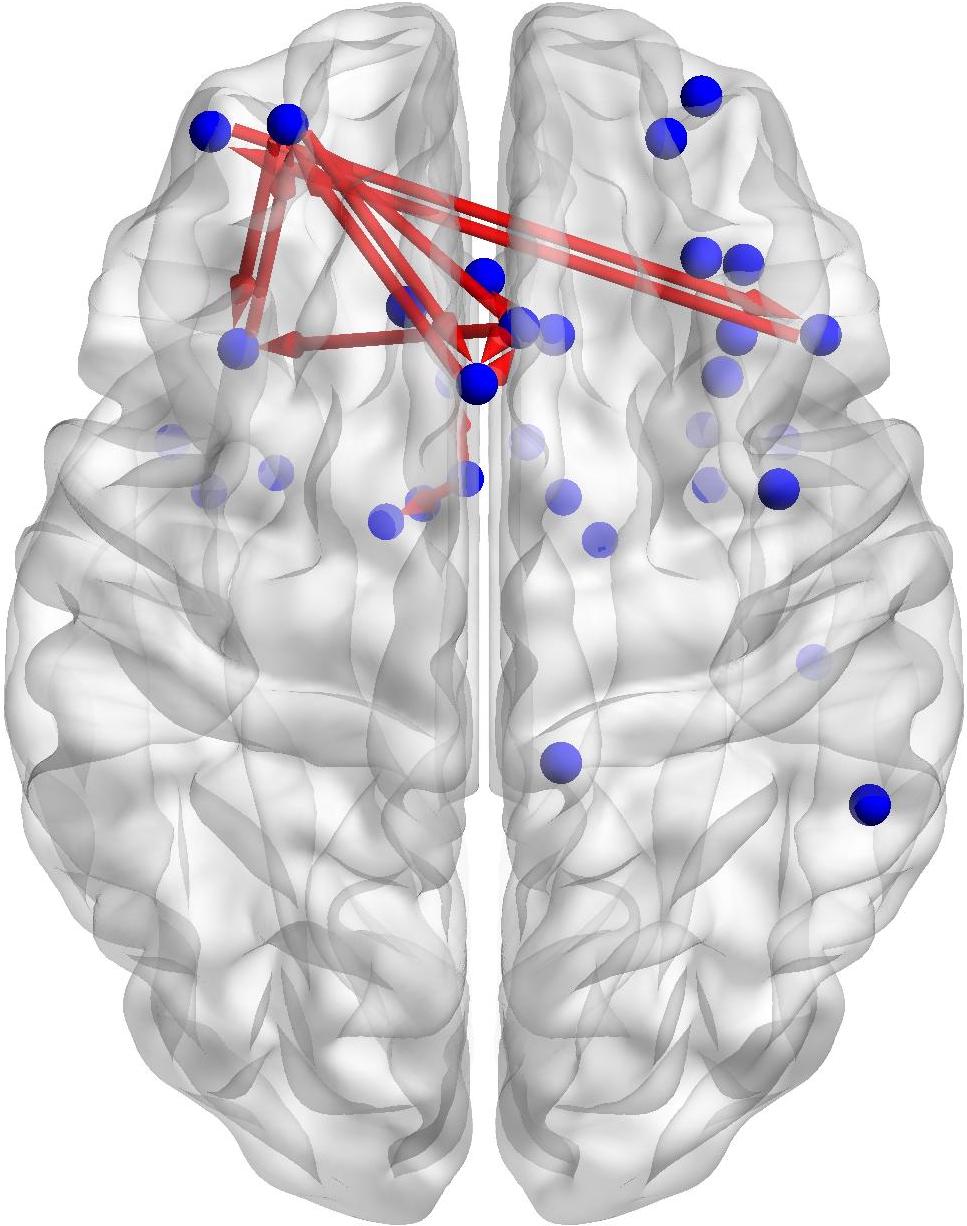} & 
&
\includegraphics[scale=0.085]{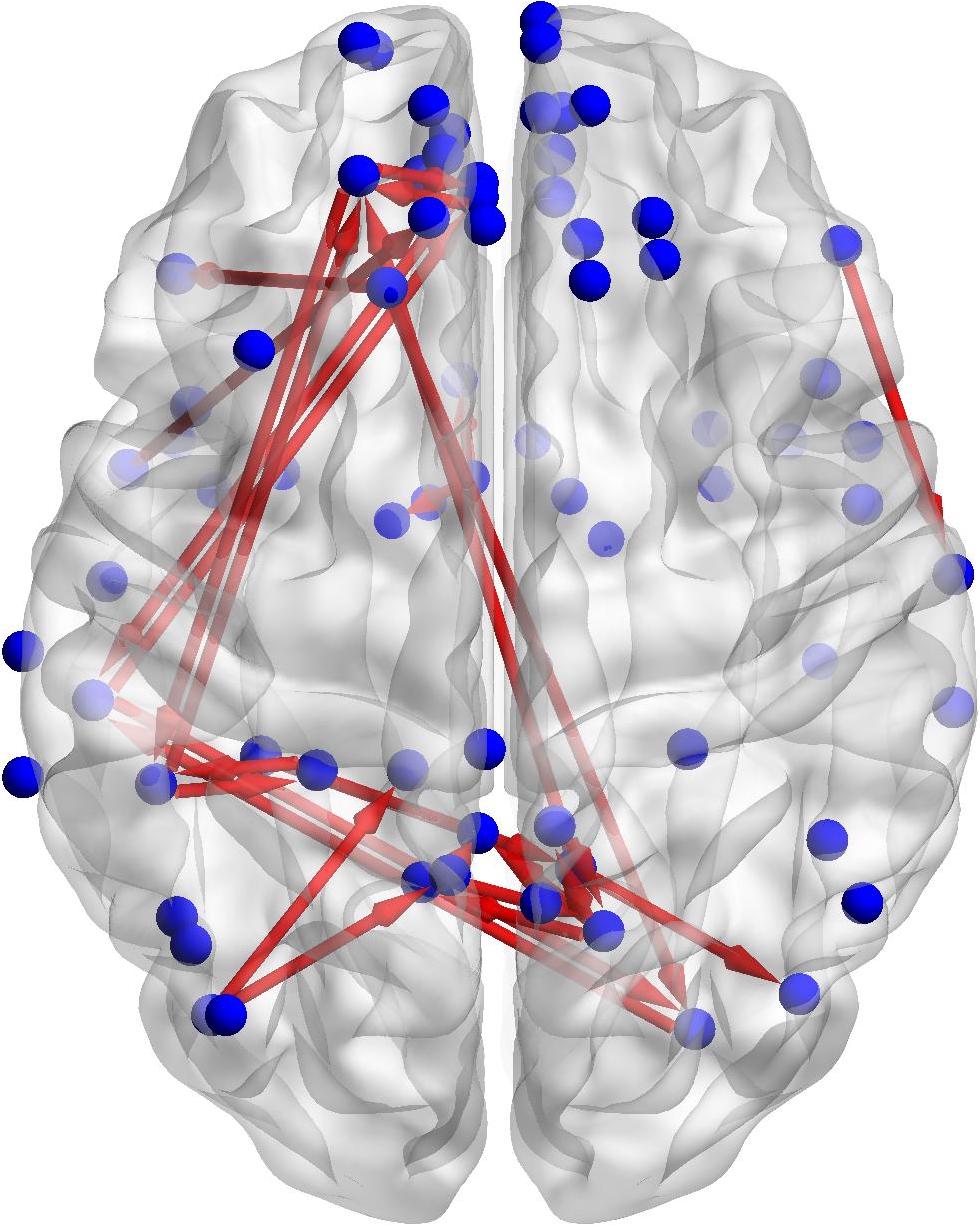} & 
\includegraphics[scale=0.085]{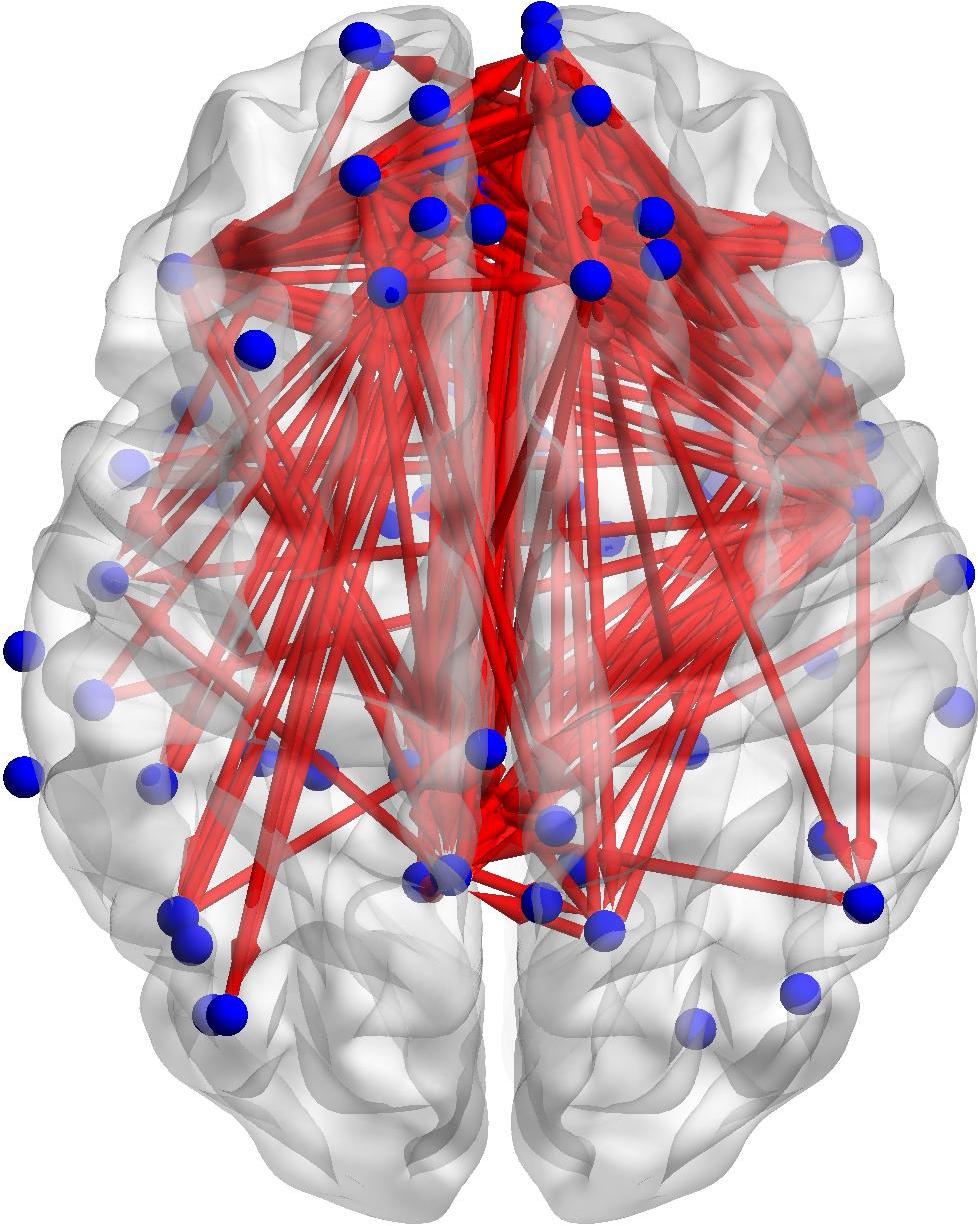} \\
high-accuracy & low-accuracy & & high-accuracy & low-accuracy \\
\\
\multicolumn{2}{c}{\textbf{Auditory}} & \hbox{     } & \multicolumn{2}{c}{\textbf{Fronto-parietal task control}} \\
\includegraphics[scale=0.085]{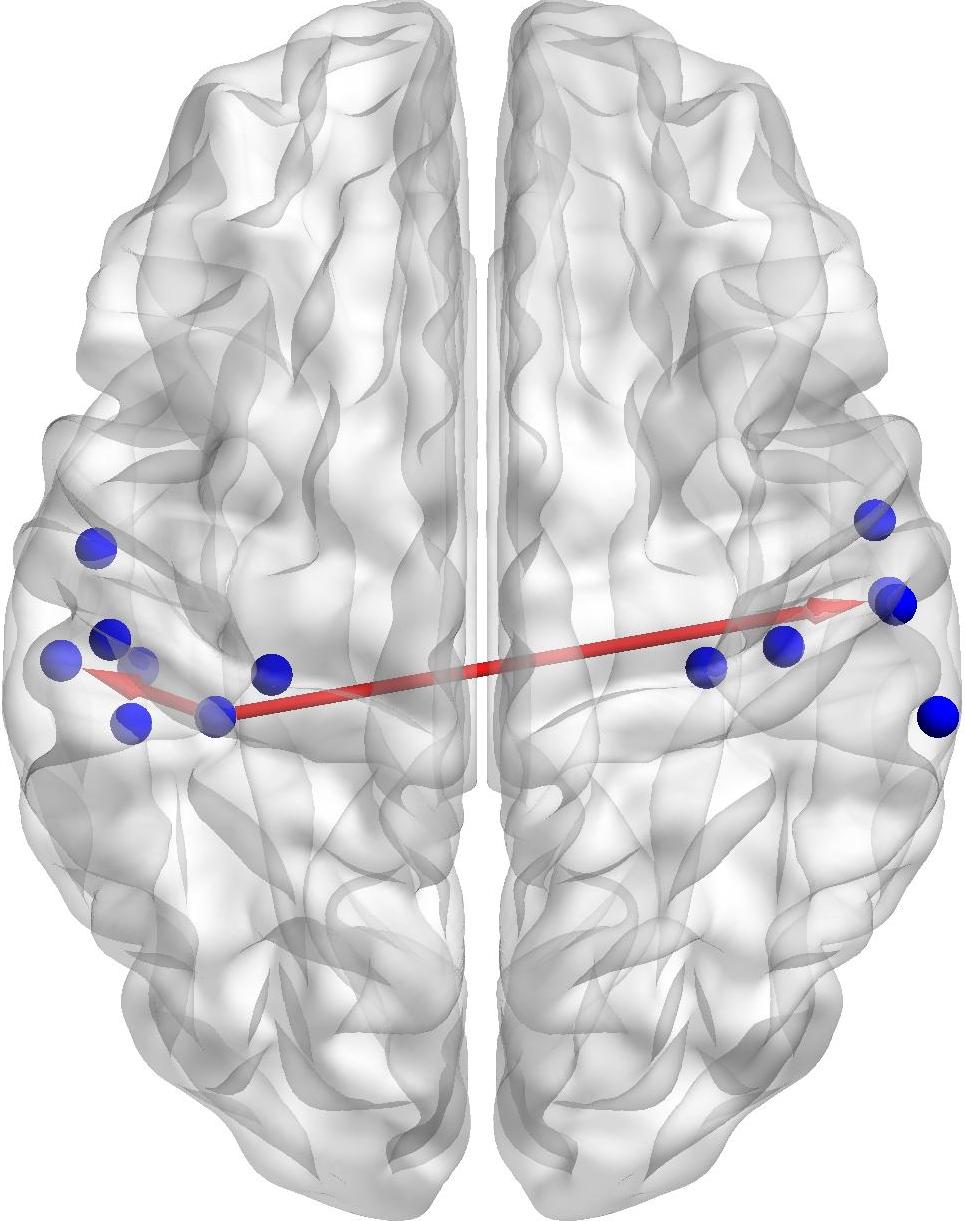} & 
\includegraphics[scale=0.085]{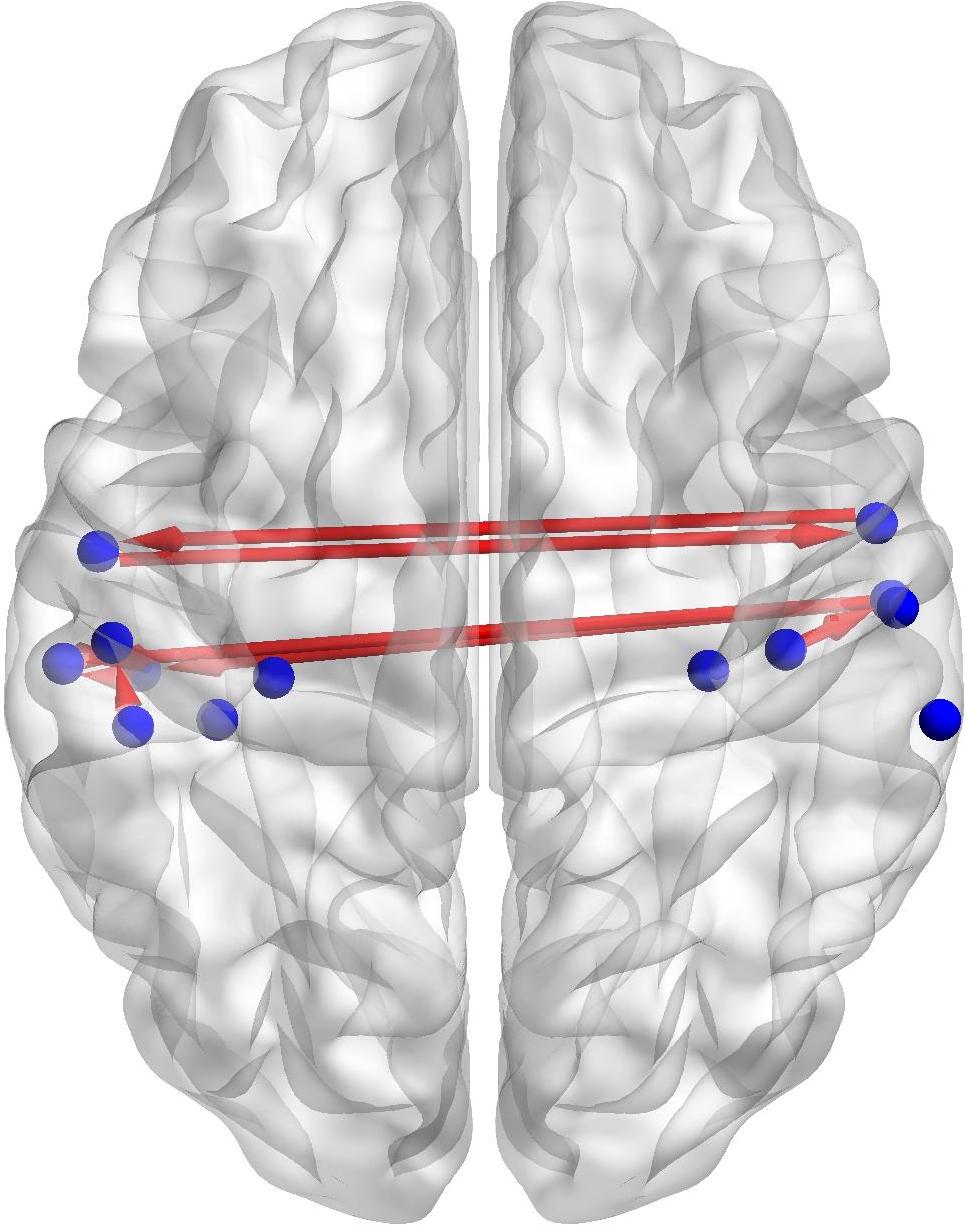} & 
&
\includegraphics[scale=0.085]{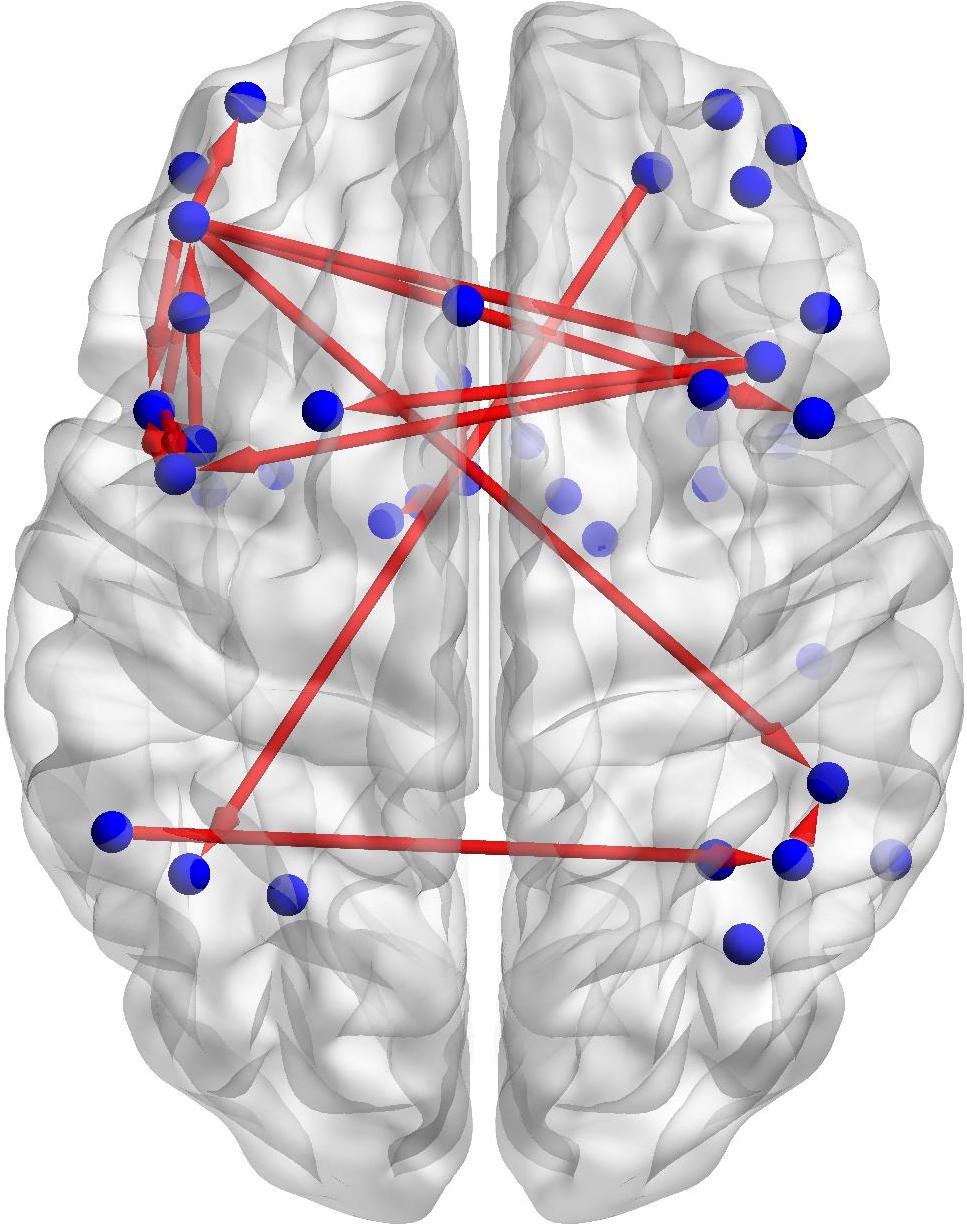} & 
\includegraphics[scale=0.085]{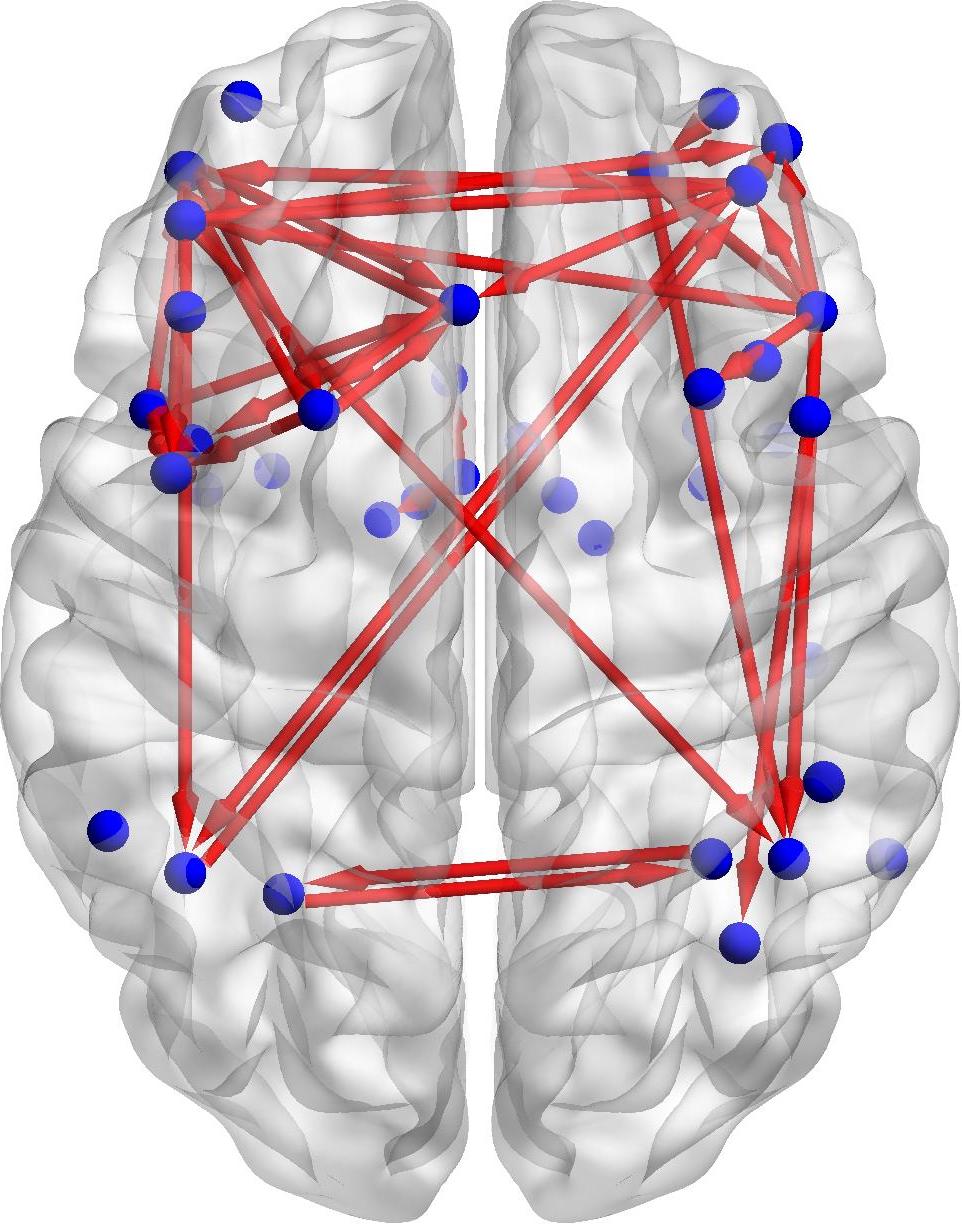} \\
high-accuracy & low-accuracy & & high-accuracy & low-accuracy \\
\end{tabular}
\caption{Visualization of the identified brain regions and their within-module connections of the high-accuracy and low-accuracy subjects for the eight functional modules.} 
\label{fig: HCP_brainnet}
\vskip 0em
\end{figure}

We begin with the global test for each subject separately. The $p$-values for the global test for both subjects are smaller than $10^{-15}$, indicating that at least one pair of brain regions have statistically significant connectivity. We then apply the simultaneous test, with the the FDR set at $0.001$. First of all, we have identified more within-module connections than the between-module connections (294 out of 7700 or $3.8\%$ versus 961 out of 61936 or $1.6\%$ for the high-accuracy subject, and 376 out of 7700 or $4.9\%$ versus 1350 out of 61936 or $2.2\%$ for the low-accuracy subject). The partition of the brain regions to the functional modules has been fully based on the biological knowledge, and our finding lends some numerical support to this partition. Second, the majority of within-module connections are concentrated on eight functional modules. Moreover, when comparing between the two subjects among those modules, we find that the high-accuracy subject has more within-module connections than the low-accuracy subject for the following functional modules: visual (118 versus 27 out of 961), salience (29 versus 11 out of 324), cingulo-opercular task control (17 versus 3 out of 196), and memory retrieval (6 versus 2 out of 25) modules. Such findings suggest that the high-accuracy subject has exhibited more intensive neural activities for processing visual imagery, memory retrieval, tonic alertness and executive control when performing the story-math task, which agrees with the literature \citep{sadaghiani2015functional, luo2014emotion}. On the other hand, we find that the high-accuracy subject has fewer connections than the low-accuracy subject for the following functional modules: default mode (25 versus 200 out of 3364), fronto-parietal task control (15 versus 37 out of 625), auditory (2 versus 8 out of 169), and subcortical (19 versus 49 out of 169) modules. These findings again agree with the literature, in that these modules have been found strongly associated with the language and reasoning type tasks \citep{schultz2016higher}, and the high-accuracy subject has exhibited less brain activity interplay related to auditory processing and mind wandering \citep{van2017mind}. Figure \ref{fig: HCP_mat} shows the identified connectivity patterns for the two subjects, and Figure \ref{fig: HCP_brainnet} shows the corresponding brain regions visualized using BrainNet Viewer \citep{Xia2013}.

\section{Discussion}
\label{sec: discussion}

In this article, we study both global and simultaneous inferences of the transition matrix under the high-dimensional vector autoregression model with measurement error. There is no existing solution, and our proposal makes a useful contribution for scientific applications such as brain connectivity analysis and others. The technical tools we develop are also of independent interest, and can facilitate the development of inferential procedures for other models involving latent variables or correlated observations. We next make some remarks regarding our model assumptions, potential limitations, and possible extensions. 

We have primarily focused on a lag-1 autoregressive structure in this article. Meanwhile, our proposal can be extended in a relatively straightforward fashion to a more general lag structure. Specifically, suppose the number of lags is $d$. Then the latent process in model \eqref{eq: model_measure} becomes $\xb_{t} = \sum_{l=1}^d \Ab_{l,*} \xb_{t-l} + \bm{\eta}_{t-1}$, and the problem of interest becomes testing $\Ab_{1,*} , \ldots , \Ab_{d,*}$.  This lag-$d$ VAR model can be equivalently rewritten as a lag-1 model, such that $\tilde{\xb}_{t} = \tilde{\Ab}_* \tilde{\xb}_{t-1} + \tilde{\bm{\eta}}_{t-1}$, $\tilde{\xb}_{t} = ({\xb}_{t}^\top, \ldots, {\xb}_{t-d+1}^\top)^\top \in \RR^{pd}$, $\tilde{\bm{\eta}}_t= (\bm{\eta}_t^\top, \zero_p^\top , \ldots , \zero_p^\top)^\top\in \RR^{pd}$, and
\begin{align*}
\tilde{\Ab}_* =
\begin{pmatrix}   
\Ab_{1,*}   &    \Ab_{2 ,*}& \ldots  &   \Ab_{d,*} \\
\Ib_p   &   \zero_{p\times p}  & \ldots &   \zero_{p\times p} \\
\zero_{p\times p}    &  \Ib_p  & \ldots &   \zero_{p\times p} \\
\zero_{p\times p}    &  \zero_{p\times p}  & \ldots &   \Ib_p 
\end{pmatrix}_{pd \times pd}.
\end{align*}
We can then apply our test to the first block row of $\tilde{\Ab}_*$, which in turn tests $\Ab_{1,*}, \ldots, \Ab_{d,*}$.

We have assumed a homoscedastic and independent error structure for both error terms $\bepsilon_t$ and $\bm{\eta}_t$. This is essentially a tradeoff. Under such an error structure, the individual variables in $\xb_t$ are still non-identically distributed and highly correlated given the autoregressive structure of the model. In applications such as brain connectivity analysis, it is often reasonable to keep a simplified error structure \citep{Zhang2015}. In the VAR literature, more general error structures have been considered. However, when estimating the transition matrix, none of existing methods directly estimated this error structure. By contrast, our inference hinges on a good estimate of the error terms. A more general form of the error structure would introduce more unknown parameters, and requires a considerable amount of extra work to characterize the estimation precision. We thus keep a simple error structure in this first work on statistical inference, and leave the more general form of the error terms as future research. 

In brain connectivity analysis, the early experiments usually focus on a single experiment subject or a very small number of subjects \citep{Friston2011}. More recently, data involving a large number of subjects are emerging. It is of interest to extend our modeling framework of a single subject to multiple subjects. The key is to capture the subject-to-subject variability by incorporating the subject-specific covariates, meanwhile integrating common information shared across different subjects. A full pursuit of this topic is beyond the scope of this article, and we leave it as future research.

\baselineskip=20pt
\bibliographystyle{ims}
\bibliography{transition_test}

\begin{thebibliography}{41}
\expandafter\ifx\csname natexlab\endcsname\relax\def\natexlab#1{#1}\fi
\expandafter\ifx\csname url\endcsname\relax
  \def\url#1{\texttt{#1}}\fi
\expandafter\ifx\csname urlprefix\endcsname\relax\def\urlprefix{URL }\fi

\bibitem[{Balakrishnan et~al.(2017)Balakrishnan, Wainwright and Yu}]{BWY17}
\textsc{Balakrishnan, S.}, \textsc{Wainwright, M.~J.} and \textsc{Yu, B.}
  (2017).
\newblock Statistical guarantees for the em algorithm: From population to
  sample-based analysis.
\newblock \textit{The Annals of Statistics} \textbf{45} 77--120.

\bibitem[{Basu and Michailidis(2015)}]{BM15}
\textsc{Basu, S.} and \textsc{Michailidis, G.} (2015).
\newblock Regularized estimation in sparse high-dimensional time series models.
\newblock \textit{The Annals of Statistics} \textbf{43} 1535--1567.

\bibitem[{Bullmore and Sporns(2009)}]{Bullmore2009}
\textsc{Bullmore, E.} and \textsc{Sporns, O.} (2009).
\newblock {Complex brain networks: graph theoretical analysis of structural and
  functional systems.}
\newblock \textit{Nature reviews. Neuroscience} \textbf{10} 186--198.

\bibitem[{Cai(2017)}]{C17}
\textsc{Cai, T.~T.} (2017).
\newblock Global testing and large-scale multiple testing for high-dimensional
  covariance structures.
\newblock \textit{Annual Review of Statistics and Its Application} \textbf{4}
  423--446.

\bibitem[{Cai and Jiang(2011)}]{CJ11}
\textsc{Cai, T.~T.} and \textsc{Jiang, T.} (2011).
\newblock Limiting laws of coherence of random matrices with applications to
  testing covariance structure and construction of compressed sensing matrices.
\newblock \textit{The Annals of Statistics} \textbf{39} 1496--1525.

\bibitem[{Cai et~al.(2013)Cai, Liu and Xia}]{Cai2013}
\textsc{Cai, T.~T.}, \textsc{Liu, W.} and \textsc{Xia, Y.} (2013).
\newblock Two-sample covariance matrix testing and support recovery in
  high-dimensional and sparse settings.
\newblock \textit{J. Amer. Statist. Assoc.} \textbf{108} 265--277.

\bibitem[{Cai et~al.(2019)Cai, Ma and Zhang}]{CMZ19}
\textsc{Cai, T.~T.}, \textsc{Ma, J.} and \textsc{Zhang, L.} (2019).
\newblock Chime: Clustering of high-dimensional gaussian mixtures with em
  algorithm and its optimality.
\newblock \textit{The Annals of Statistics} \textbf{47} 1234--1267.

\bibitem[{Cai and Sun(2017)}]{CaiSun2017}
\textsc{Cai, T.~T.} and \textsc{Sun, W.} (2017).
\newblock Large-scale global and simultaneous inference: Estimation and testing
  in very high dimensions.
\newblock \textit{Annual Review of Economics} \textbf{9} 411--439.

\bibitem[{Candes and Tao(2007)}]{CT07}
\textsc{Candes, E.} and \textsc{Tao, T.} (2007).
\newblock The dantzig selector: Statistical estimation when p is much larger
  than n.
\newblock \textit{The Annals of Statistics} \textbf{35} 2313--2351.

\bibitem[{Cao et~al.(2019)Cao, Sandstede and Luo}]{Luo2019}
\textsc{Cao, X.}, \textsc{Sandstede, B.} and \textsc{Luo, X.} (2019).
\newblock A functional data method for causal dynamic network modeling of
  task-related fmri.
\newblock \textit{Frontiers in Neuroscience} \textbf{13} 127.

\bibitem[{Chen et~al.(2011)Chen, Glen, Saad, Hamilton, Thomason, Gotlib and
  Cox}]{Chen2011}
\textsc{Chen, G.}, \textsc{Glen, D.}, \textsc{Saad, Z.}, \textsc{Hamilton,
  J.~P.}, \textsc{Thomason, M.}, \textsc{Gotlib, I.} and \textsc{Cox, R.}
  (2011).
\newblock Vector autoregression, structural equation modeling, and their
  synthesis in neuroimaging data analysis.
\newblock \textit{Computers in biology and medicine} \textbf{41} 1142--55.

\bibitem[{Chen et~al.(2010)Chen, Zhang and Zhong}]{CZZ10}
\textsc{Chen, S.~X.}, \textsc{Zhang, L.-X.} and \textsc{Zhong, P.-S.} (2010).
\newblock Tests for high-dimensional covariance matrices.
\newblock \textit{Journal of the American Statistical Association} \textbf{105}
  810--819.

\bibitem[{Friston(2011)}]{Friston2011}
\textsc{Friston, K.~J.} (2011).
\newblock Functional and effective connectivity: A review.
\newblock \textit{Brain Connectivity} \textbf{1} 13--36.

\bibitem[{Garg et~al.(2011)Garg, Cecchi and Rao}]{Garg2011}
\textsc{Garg, R.}, \textsc{Cecchi, G.} and \textsc{Rao, R.} (2011).
\newblock Full-brain auto-regressive modeling (farm) using fmri.
\newblock \textit{NeuroImage} \textbf{58} 416--41.

\bibitem[{Ghahramani and Hinton(1996)}]{GH96}
\textsc{Ghahramani, Z.} and \textsc{Hinton, G.~E.} (1996).
\newblock Parameter estimation for linear dynamical systems.
\newblock Tech. rep., University of Toronto.

\bibitem[{Glasser et~al.(2013)Glasser, Sotiropoulos, Wilson, Coalson, Fischl,
  Andersson, Xu, Jbabdi, Webster, Polimeni et~al.}]{glasser2013minimal}
\textsc{Glasser, M.~F.}, \textsc{Sotiropoulos, S.~N.}, \textsc{Wilson, J.~A.},
  \textsc{Coalson, T.~S.}, \textsc{Fischl, B.}, \textsc{Andersson, J.~L.},
  \textsc{Xu, J.}, \textsc{Jbabdi, S.}, \textsc{Webster, M.}, \textsc{Polimeni,
  J.~R.} \textsc{et~al.} (2013).
\newblock The minimal preprocessing pipelines for the human connectome project.
\newblock \textit{Neuroimage} \textbf{80} 105--124.

\bibitem[{Han et~al.(2015)Han, Lu and Liu}]{HLL15}
\textsc{Han, F.}, \textsc{Lu, H.} and \textsc{Liu, H.} (2015).
\newblock A direct estimation of high dimensional stationary vector
  autoregressions.
\newblock \textit{The Journal of Machine Learning Research} \textbf{16}
  3115--3150.

\bibitem[{Hsu et~al.(2008)Hsu, Hung and Chang}]{HHC08}
\textsc{Hsu, N.-J.}, \textsc{Hung, H.-L.} and \textsc{Chang, Y.-M.} (2008).
\newblock Subset selection for vector autoregressive processes using lasso.
\newblock \textit{Computational Statistics \& Data Analysis} \textbf{52}
  3645--3657.

\bibitem[{Krampe et~al.(2018)Krampe, Kreiss and Paparoditis}]{KKP18}
\textsc{Krampe, J.}, \textsc{Kreiss, J.} and \textsc{Paparoditis, E.} (2018).
\newblock Bootstrap based inference for sparse high-dimensional time series
  models.
\newblock \textit{arXiv preprint arXiv:1806.11083} .

\bibitem[{Liu(2013)}]{L13}
\textsc{Liu, W.} (2013).
\newblock Gaussian graphical model estimation with false discovery rate
  control.
\newblock \textit{The Annals of Statistics} \textbf{41} 2948--2978.

\bibitem[{Liu and Shao(2014)}]{LS14}
\textsc{Liu, W.} and \textsc{Shao, Q.-M.} (2014).
\newblock Phase transition and regularized bootstrap in large-scale $ t $-tests
  with false discovery rate control.
\newblock \textit{The Annals of Statistics} \textbf{42} 2003--2025.

\bibitem[{Luo et~al.(2014)Luo, Qin, Fernandez, Zhang, Klumpers and
  Li}]{luo2014emotion}
\textsc{Luo, Y.}, \textsc{Qin, S.}, \textsc{Fernandez, G.}, \textsc{Zhang, Y.},
  \textsc{Klumpers, F.} and \textsc{Li, H.} (2014).
\newblock Emotion perception and executive control interact in the salience
  network during emotionally charged working memory processing.
\newblock \textit{Human Brain Mapping} \textbf{35} 5606--5616.

\bibitem[{Negahban and Wainwright(2011)}]{NW11}
\textsc{Negahban, S.} and \textsc{Wainwright, M.~J.} (2011).
\newblock Estimation of (near) low-rank matrices with noise and
  high-dimensional scaling.
\newblock \textit{The Annals of Statistics} \textbf{39} 1069--1097.

\bibitem[{Ning and Liu(2017)}]{NL17}
\textsc{Ning, Y.} and \textsc{Liu, H.} (2017).
\newblock A general theory of hypothesis tests and confidence regions for
  sparse high dimensional models.
\newblock \textit{The Annals of Statistics} \textbf{45} 158--195.

\bibitem[{Power et~al.(2011)Power, Cohen, Nelson, Wig, Barnes, Church, Vogel,
  Laumann, Miezin, Schlaggar et~al.}]{power2011functional}
\textsc{Power, J.~D.}, \textsc{Cohen, A.~L.}, \textsc{Nelson, S.~M.},
  \textsc{Wig, G.~S.}, \textsc{Barnes, K.~A.}, \textsc{Church, J.~A.},
  \textsc{Vogel, A.~C.}, \textsc{Laumann, T.~O.}, \textsc{Miezin, F.~M.},
  \textsc{Schlaggar, B.~L.} \textsc{et~al.} (2011).
\newblock Functional network organization of the human brain.
\newblock \textit{Neuron} \textbf{72} 665--678.

\bibitem[{Reinsel(2003)}]{Reinsel_book}
\textsc{Reinsel, G.~C.} (2003).
\newblock \textit{Elements of multivariate time series analysis}.
\newblock Springer Science \& Business Media.

\bibitem[{Sadaghiani and D'Esposito(2015)}]{sadaghiani2015functional}
\textsc{Sadaghiani, S.} and \textsc{D'Esposito, M.} (2015).
\newblock Functional characterization of the cingulo-opercular network in the
  maintenance of tonic alertness.
\newblock \textit{Cerebral Cortex} \textbf{25} 2763--2773.

\bibitem[{Schultz and Cole(2016)}]{schultz2016higher}
\textsc{Schultz, D.~H.} and \textsc{Cole, M.~W.} (2016).
\newblock Higher intelligence is associated with less task-related brain
  network reconfiguration.
\newblock \textit{Journal of neuroscience} \textbf{36} 8551--8561.

\bibitem[{Shao(2015)}]{S15}
\textsc{Shao, X.} (2015).
\newblock Self-normalization for time series: a review of recent developments.
\newblock \textit{Journal of the American Statistical Association} \textbf{110}
  1797--1817.

\bibitem[{Smith et~al.(2009)Smith, Fox, Miller, Glahn, Fox, Mackay, Filippini,
  Watkins, Toro, Laird and Beckmann}]{Smith2009}
\textsc{Smith, S.~D.}, \textsc{Fox, P.~T.}, \textsc{Miller, K.}, \textsc{Glahn,
  D.}, \textsc{Fox, P.}, \textsc{Mackay, C.~E.}, \textsc{Filippini, N.},
  \textsc{Watkins, K.~E.}, \textsc{Toro, R.}, \textsc{Laird, A.} and
  \textsc{Beckmann, C.~F.} (2009).
\newblock Correspondence of the brain; functional architecture during
  activation and rest.
\newblock \textit{Proceedings of the National Academy of Sciences of the United
  States of America} \textbf{106} 13040--5.

\bibitem[{Song and Bickel(2011)}]{SB11}
\textsc{Song, S.} and \textsc{Bickel, P.~J.} (2011).
\newblock Large vector auto regressions.
\newblock \textit{arXiv preprint arXiv:1106.3915} .

\bibitem[{Tsay and Chen(2018)}]{tsay_book}
\textsc{Tsay, R.~S.} and \textsc{Chen, R.} (2018).
\newblock \textit{Nonlinear time series analysis}, vol. 891.
\newblock John Wiley \& Sons.

\bibitem[{Van~Essen et~al.(2013)Van~Essen, Smith, Barch, Behrens, Yacoub,
  Ugurbil, Consortium et~al.}]{van2013wu}
\textsc{Van~Essen, D.~C.}, \textsc{Smith, S.~M.}, \textsc{Barch, D.~M.},
  \textsc{Behrens, T.~E.}, \textsc{Yacoub, E.}, \textsc{Ugurbil, K.},
  \textsc{Consortium, W.-M.~H.} \textsc{et~al.} (2013).
\newblock The wu-minn human connectome project: an overview.
\newblock \textit{Neuroimage} \textbf{80} 62--79.

\bibitem[{Van~Praag et~al.(2017)Van~Praag, Garfinkel, Sparasci, Mees,
  Philippides, Ware, Ottaviani and Critchley}]{van2017mind}
\textsc{Van~Praag, C. D.~G.}, \textsc{Garfinkel, S.~N.}, \textsc{Sparasci, O.},
  \textsc{Mees, A.}, \textsc{Philippides, A.~O.}, \textsc{Ware, M.},
  \textsc{Ottaviani, C.} and \textsc{Critchley, H.~D.} (2017).
\newblock Mind-wandering and alterations to default mode network connectivity
  when listening to naturalistic versus artificial sounds.
\newblock \textit{Scientific Reports} \textbf{7} 45273.

\bibitem[{Wang et~al.(2015)Wang, Gu, Ning and Liu}]{WGNL15}
\textsc{Wang, Z.}, \textsc{Gu, Q.}, \textsc{Ning, Y.} and \textsc{Liu, H.}
  (2015).
\newblock High dimensional em algorithm: Statistical optimization and
  asymptotic normality.
\newblock In \textit{Advances in neural information processing systems}.

\bibitem[{Xia et~al.(2013)Xia, Wang and He}]{Xia2013}
\textsc{Xia, M.}, \textsc{Wang, J.} and \textsc{He, Y.} (2013).
\newblock Brainnet viewer: A network visualization tool for human brain
  connectomics.
\newblock \textit{PLOS ONE} \textbf{8} 1--15.

\bibitem[{Xia et~al.(2018)Xia, Cai and Cai}]{XCC18}
\textsc{Xia, Y.}, \textsc{Cai, T.} and \textsc{Cai, T.~T.} (2018).
\newblock Multiple testing of submatrices of a precision matrix with
  applications to identification of between pathway interactions.
\newblock \textit{Journal of the American Statistical Association} \textbf{113}
  328--339.

\bibitem[{Xiao and Wu(2013)}]{XW13}
\textsc{Xiao, H.} and \textsc{Wu, W.~B.} (2013).
\newblock Asymptotic theory for maximum deviations of sample covariance matrix
  estimates.
\newblock \textit{Stochastic Processes and their Applications} \textbf{123}
  2899--2920.

\bibitem[{Yi and Caramanis(2015)}]{YC15}
\textsc{Yi, X.} and \textsc{Caramanis, C.} (2015).
\newblock Regularized em algorithms: A unified framework and statistical
  guarantees.
\newblock In \textit{Advances in Neural Information Processing Systems}.

\bibitem[{Zhang et~al.(2015)Zhang, Wu, Li, Caffo and Boatman-Reich}]{Zhang2015}
\textsc{Zhang, T.}, \textsc{Wu, J.}, \textsc{Li, F.}, \textsc{Caffo, B.} and
  \textsc{Boatman-Reich, D.} (2015).
\newblock A dynamic directional model for effective brain connectivity using
  electrocorticographic {(ECoG)} time series.
\newblock \textit{Journal of the American Statistical Association} \textbf{110}
  93--106.

\bibitem[{Zheng and Raskutti(2019)}]{ZC19}
\textsc{Zheng, L.} and \textsc{Raskutti, G.} (2019).
\newblock Testing for high-dimensional network parameters in auto-regressive
  models.
\newblock \textit{Electronic Journal of Statistics} \textbf{13} 4977--5043.

\end{thebibliography}

\end{document}